\documentclass[useAMS,usenatbib]{mn2e}

\usepackage[dvips]{graphicx}

\def\kms{{\rm km}\;{\rm s}^{-1}}

\title[Accurate fundamental parameters for Lower Main Sequence Stars]
{Accurate fundamental parameters for Lower Main Sequence Stars.}

\author[Casagrande, Portinari and Flynn]{Luca Casagrande$^{1}$\thanks{E-mail: 
luccas@utu.fi (LC), lporti@utu.fi (LP), cflynn@astro.utu.fi (CF)}, Laura 
Portinari$^{1}$ and Chris Flynn$^{1,2}$\\
$^1$ Tuorla Observatory, V\"ais\"al\"antie 20, FI-21500 Piikki\"o, Finland\\
$^2$ Mount Stromlo Observatory, Cotter Road, Weston, ACT, Australia}

\begin{document}

\maketitle

\begin{abstract}

We derive an empirical effective temperature and bolometric luminosity
calibration for G and K dwarfs, by applying our own implementation of the
InfraRed Flux Method to multi--band photometry. Our study is based on 104 stars
for which we have excellent $BV(RI)_C JHK_S$ photometry, excellent parallaxes
and good metallicities.

Colours computed from the most recent synthetic libraries (ATLAS9 and MARCS)
are found to be in good agreement with the empirical colours in the optical
bands, but some discrepancies still remain in the infrared. Synthetic and
empirical bolometric corrections also show fair agreement. 

A careful comparison to temperatures, luminosities and angular diameters
obtained with other methods in literature shows that systematic effects still
exist in the calibrations at the level of a few percent. Our InfraRed Flux
Method temperature scale is 100~K hotter than recent analogous determinations
in the literature, but is in agreement with spectroscopically calibrated
temperature scales and fits well the colours of the Sun.  Our angular diameters
are typically 3\% smaller when compared to other (indirect) determinations of
angular diameter for such stars, but are consistent with the limb-darkening
corrected predictions of the latest 3D model atmospheres and also with the
results of asteroseismology.

Very tight empirical relations are derived for bolometric luminosity, effective
temperature and angular diameter from photometric indices.  

We find that much of the discrepancy with other temperature scales and the
uncertainties in the infrared synthetic colours arise from the uncertainties in
the use of Vega as the flux calibrator.  Angular diameter measurements for a
well chosen set of G and K dwarfs would go a long way to addressing this
problem.

\end{abstract}


\section{Introduction}
\label{intro}

Temperatures, luminosities and radii are amongst the basic physical data
against which models of stellar structure and evolution are tested.  In this
paper, we address one particular area of the Hertzsprung-Russell (HR) diagram,
that of the G and K dwarfs --- focusing particularly on stars for which the
effects of stellar evolution have been negligible or nearly negligible during
their lifetimes. Somewhat higher mass stars than those considered here have
been extensively studied historically because their luminosities may be used to
infer stellar ages. Lower mass stars of late-G to K spectral types, have been
neglected to some extent, probably because there have been few secondary
benefits in getting stellar models right for these stars, and the lack of good
parallax, diameter and other data. This situation is rapidly changing: firstly,
Hipparcos has provided the requisite parallax data; secondly, interferometric
techniques are making the measurement of diameters for such small stars likely
to be routine within a few years; thirdly, we have our own motivation in the
form of an ongoing project to follow the chemical history of the Milky Way from
lower mass stars, for which we can infer indirectly the amount of their
constituent helium via stellar luminosity (Jimenez et al 2003). In order to
achieve our goals, we need accurate and homogeneous effective temperatures and
luminosities for our G+K dwarfs.

The effective temperature of a stellar surface is a measure of the total
energy, integrated over all wavelengths, radiated from a unit of surface
area. Since its value is fixed by the luminosity and radius, it is readily
calculated for theoretical stellar models, and as one of the coordinates of the
physical HR diagram, it plays a central role in discussions of stellar
evolution. Most observations, however, provide spectroscopic or photometric
indicators of temperature that are only indirectly related to the effective
temperature.

The effective temperature of lower main sequence stars is not easy to determine
and different measurement techniques are still far from satisfactory
concordance (e.g. Mishenina \& Kovtyukh 2001, Kovtyukh et al. 2003). At
present, spectroscopic temperature determinations return values that are some
100~K hotter that most of the other techniques.  Even in a restricted and
thoroughly studied region like that of the Solar Analogs, effective
temperature determinations for the same star still differ significantly, by as
much as 150~K (Soubiran \& Triaud, 2004). Models predict values that are some
100~K hotter than those measured (Lebreton et al. 1999 and references therein).

In this work, we apply the Infrared Flux Method to multi-band $BV(RI)_{C}JHK_S$
photometry of a carefully selected sample of G and K dwarf stars.  We compare
observed with synthetic broad-band colours computed from up-to-date (1D) model
atmospheres. Such models are then used to estimate the missing flux needed to
recover bolometric luminosities from our photometry. Other than the high
quality of the observational data, the strength of this work relies on using
very few basic assumptions: these are the adopted Vega absolute calibration and
zero-points. This also makes clearer the evaluation of possible errors
and/or biases in the results.  Both the absolute calibration and the 
zero-points
are expected to be well known and the latest generations of model atmospheres
produce realistic fluxes for a wide range of temperatures, gravities and
abundances (Bessell 2005) so that the adopted model and calibration are at the 
best level currently available.

The paper is organized as follows: in Section 2 we describe our sample; we
compare different libraries of model atmospheres with observational data in 
Section 3 and in Section 4 we present our implementation of the Infrared Flux
Method along with the resulting temperature scale. In Section 5 we test our
scale with empirical data for the Sun and solar analogs and with recent
interferometric measurements of dwarf stars.
The comparison with other temperature determinations is done in Section
6 and in Section 7 and 8 we give bolometric calibrations and useful tight
relations between angular diameters and photometric indices. We conclude in 
Section 9. 

\section{Sample selection and observations}
\label{sample}

\begin{figure*}
\begin{center}
\includegraphics[scale=0.40]{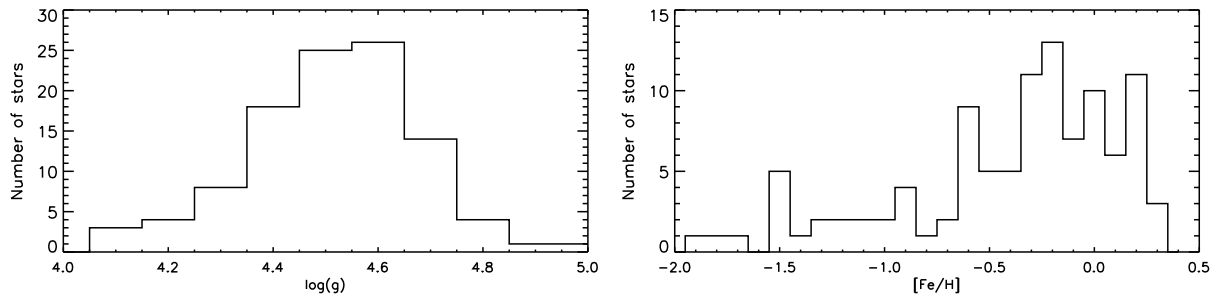}
\caption{{Distribution of $\log (g)$ and [Fe/H] for the 104 dwarf stars in the
basic sample, collected from the literature. As noted in Section \ref{libraries}
and \ref{Felog}, adoption of an average $\log(g)$ of 4.5 for all the stars is
sufficiently accurate for our purposes. All stars but one (HD 108564) with
$\textrm{[Fe/H]} < -1$ have direct $\alpha$--element measurements.}}
\label{logg}
\end{center}
\end{figure*}

The bulk of the targets has been selected from the sample of nearby stars in
the northern hemisphere provided by Gray et al. (2003).  Our initial sample
consisted of 186 G and K dwarfs with Hipparcos parallaxes for which the
relative error is less than 14\%.  Accurate $BV(RI)_{C}$ broadband photometry
has been obtained for the bulk of the stars at La Palma, with additional
photometry adopted from Bessell (1990a), Reid et al. (2001) and Percival,
Salaris \& Kilkenny (2003). All our stars have accurate spectroscopic
metallicities and for most of them also $\alpha$--element abundances.  Our
sample is neither volume nor magnitude limited, but it gives a good coverage of
the properties of local G and K dwarfs from low to high metallicity (Figure
\ref{logg}). The sample has been cleaned of a number of contaminants, reducing
it to a final sample of 104 stars, as described in the following sections.

\subsection{Removing double/multiple and variable stars}
\label{double}

Particular attention has been paid to removing unresolved double/multiple
stars. These stars primarily affect spectroscopic and photometric measurements,
making the system appear brighter and redder. The Hipparcos catalogue was used
to make a prior exclusion of 'certain', 'possible' and 'suspected' (i.e.
\emph{good, fair, poor, uncertain} and \emph{suspected}) multiple systems, on
the basis of the quality of the MultFlag field in the Hipparcos catalogue.

Some unrecognized double/multiple stars almost certainly remain in the sample,
since a few stars are found in a 'binary main sequence' above the bulk of the
main sequence, as Kotoneva, Flynn \& Jimenez (2002) suspected in such data but
were unable to prove. A new technique for identifying binaries/multiples has
been introduced by Wielen et al. (1999), termed the $\Delta\mu$ method, which
indicates the presence of multiple unresolved systems by using the difference
between the near-instantaneous direction of the proper motion of the star(s) as
measured by the Hipparcos satellite, and the direction of the ground based
proper motion measured over much longer time scales. Most of the suspect stars
in the Kotoneva, Flynn \& Jimenez (2002) sample turned out to be $\Delta\mu$
multiples, and we have excluded those stars which are likely $\Delta\mu$
multiples also from the present sample ($\Delta\mu$ classifications were kindly
made for our sample stars by Christian Dettbarn in Heidelberg). This reduced
the sample to 134 stars.

We have retained widely separated binary systems in the sample where components
could be studied separately. These were retained by making a prior
cross-checking of our sample with the catalogue of wide-binary and multiple
systems of Poveda et al. (1994) and then verifying the separation of such
systems by inspecting our own or SIMBAD images. 13 such stars have been
retained.

Even if dwarfs stars are not expected to show signs of strong variability, the
existence of $\sim 100$ or more Hipparcos measurements per star spread over
several years, together with the excellent temporal stability of the magnitude
scales, allows the detection of variability at the level of few hundredths of a
magnitude (Brown et al. 1997) in Hipparcos targets. We use the Hipparcos
classification to remove so-called 'duplicity induced variables', 'periodic
variables', 'unsolved variables' and 'possible micro-variables'.  This removed
a total of 26 stars (however 8 of these were already removed as non-single
stars).

The final number of stars satisfying the above requirements was 129.

\subsection{Broad-Band Photometric Observations}
\label{photo}

To recover accurate bolometric fluxes and temperatures for the stars, we have
obtained accurate and homogeneous Johnson-Cousins $BV(RI)_{C}$ and $JHK_{S}$
photometry for all the 186 stars in our initial sample.

\subsubsection{Johnson-Cousins photometry}
\label{bvriphoto}

For most of the stars in our sample with declination north of $\delta =
-25^{\circ}$, we have made our own photometric observations from April to
December 2004. Observations were done from Finland in full remote mode, using
the 35-cm telescope piggybacked on the Swedish 60-cm telescope located at La
Palma in the Canary Islands. An SBIG charge-coupled device was used through all
the observations. Johnson-Cousins $BV(RI)_{C}$ colours were obtained for all
stars.

Standard stars were selected from Landolt (1992) and among E-regions from
Cousins (1980). Although E-regions provide an extremely accurate set of stars,
there are some systematic differences between the Landolt and the SAAO system
(Bessell, 1995). Therefore, we have used the Landolt (1992)
standards placed onto the SAAO system by Bessell (1995).

Ten to twenty standard stars were observed each night, bracketing our program
stars in colour and airmass. Whenever possible program stars were observed when
passing the meridian, in order to minimize extrapolation to zero-airmass.  Only
if the standard deviation between our calibrated and the tabulated values for
the standards was smaller than 0.015 mag in each band was the night was
considered photometric and observations useful.  In addition, only program
stars for which the final $B-V$ scatter (obtained averaging five frames) was
smaller than 0.015 mag were considered usefully accurate. Scatter in the other
bands was usually smaller. We expect our photometry to have accuracies of
0.010-0.015 mag on average.

In addition to our observations, we have gathered $BV(RI)_{C}$ photometry of
equivalent or better precision from the literature : Bessell (1990a) (who also
include measurements from Cousins (1980)), Reid et al. (2001) and Percival,
Salaris \& Kilkenny (2003). For stars in common with these authors, we have
found excellent agreement between the photometry, with a scatter of the
order of 0.01 mag in all bands, and zero-point shifts between authors of 
less than 0.01 mag. This is more than adequately accurate for our study. 

\subsubsection{2MASS photometry}

Infrared $JHK_{S}$ photometry for the sample has been taken from the 2MASS
catalogue. The uncertainty for each observed 2MASS magnitude is denoted in the
catalogue by the flags : flags ``j\_'', ``h\_'' and ``k\_msigcom'', and is the
complete error that incorporates the results of processing photometry, internal
errors and calibration errors. Some of our stars are very bright and have very
high errors in 2MASS. We use 2MASS photometry only if the sum of the
photometric errors is less than 0.10 mag (i.e. ``j\_''$+$``h\_''$+$
``k\_msigcom''$< 0.10$). The final number of stars is thus reduced from 129 to
104. For our final sample the errors in $J$ and $K_{S}$ bands are similar, with
a mean value of 0.02 mag, whereas a slightly higher mean error is found in $H$
band (0.03 mag).

\begin{table*}
 \vbox to220mm{\vfill Landscape table to go here.
 \caption{}
 \vfill}
 \label{stars}
\end{table*}

\subsection{Abundances}

We have gathered detailed chemical abundances for our sample stars from the
wealth of on-going surveys, dedicated to investigate the chemical composition
of our local environment as well as to the host stars of extra-solar planets
(see references in Table 1).  The internal accuracy of such data is usually
excellent, with uncertainties in the order of 0.10 dex or less. However,
abundance analysis for late-type dwarfs can still be troublesome in some cases
(e.g. Allende Prieto et al. 2004).  We are aware that, in gathering
spectroscopy from different authors, the underlying temperature scale used to
derive abundances can differ by as much as 50-150~K, which would translate into
a [Fe/H] error of $\sim 0.1$ dex (e.g. Asplund 2003; Kovtyukh et
al. 2003). Therefore, a more conservative error estimate for our abundances
would be $\sim 0.15$ dex. However, as we show in Section \ref{Felog} the
InfraRed Flux Method, which we employ to recover the fundamental stellar
parameters, is only weakly sensitive to the adopted metallicity and
uncertainties of $\pm 0.15$~dex do not bear heavily on our main results.

The best measured elemental abundance in our dwarf stars is usually iron
(i.e. [Fe/H]) whereas for theoretical models the main metallicity parameter is
the total heavy-element mass fraction, [M/H].  For most of the stars in our
sample, the spectra provide measurement not only for [Fe/H], but also for the
$\alpha$-elements, which dominate the global metallicity budget. For stars with
$\alpha$-element estimates, we compute [M/H] (Yi et al., 2001):

\begin{equation}
\label{overme}
   \textrm{[M/H]}=\textrm{[Fe/H]}+\log(0.694f_{\alpha}+0.306)
\end{equation}

where 

\begin{equation}
   f_{\alpha}=10^{[\alpha/{\mathrm Fe}]}
\end{equation}

is the enhancement factor and [$\alpha$/Fe] has been computed by averaging the
$\alpha$-elements. The older formula by Salaris et al. (1993) give similar
results with a difference smaller than 0.02 dex in [M/H] even for the most
$\alpha$-enhanced stars ([$\alpha$/Fe] $\sim 0.4$).

For stars for which the $\alpha$ elements were not available we have estimated
their contribution from the mean locus of the [$\alpha$/Fe] vs. [Fe/H] relation
from the analytical model of Pagel \& Tautvai\v sien\. e (1995). There were 34 
such stars in our final sample compared to 70 for which $\alpha$ estimates 
were directly available.

\begin{figure*}
\begin{center}
\includegraphics[scale=0.86]{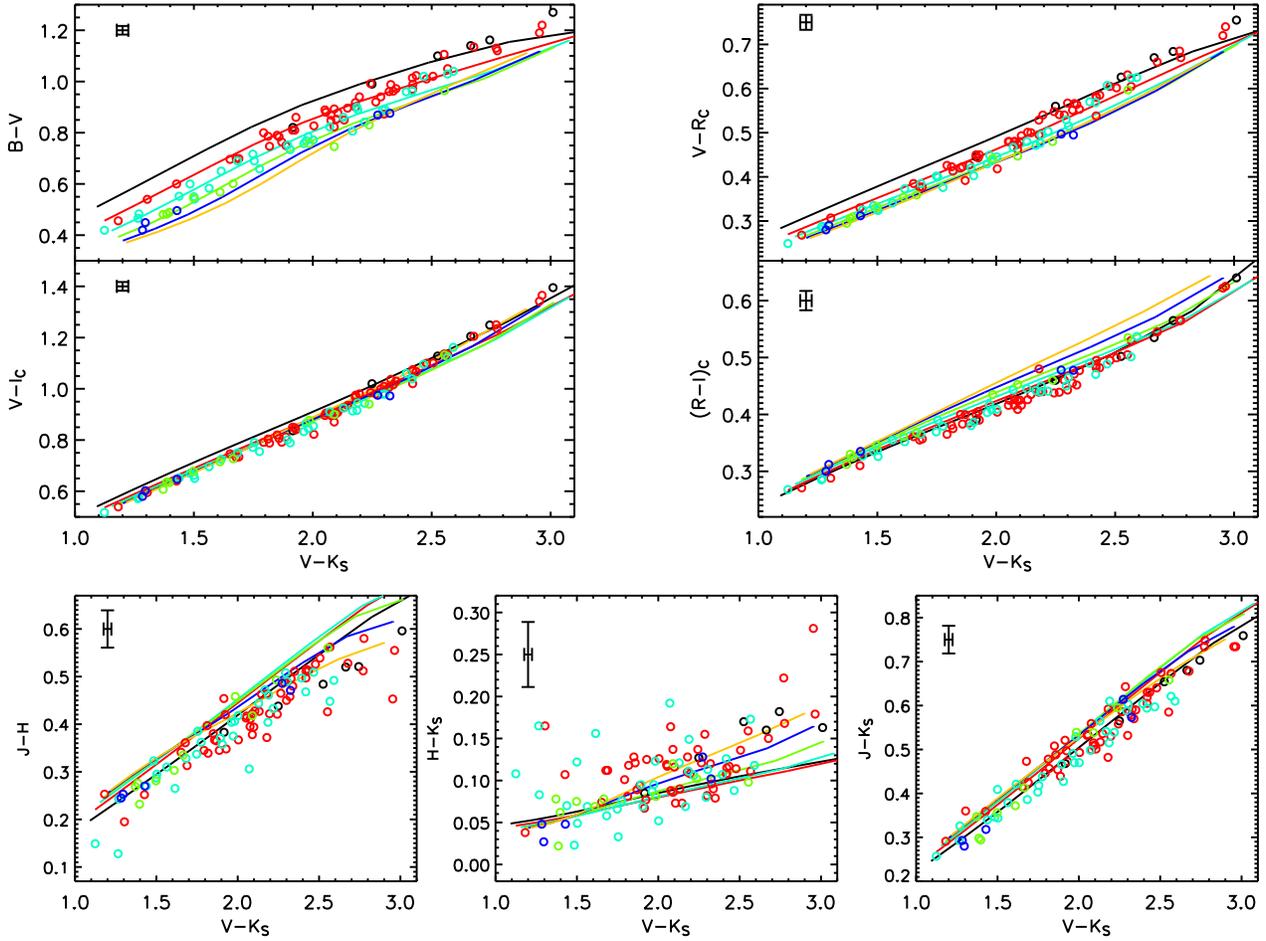}
\caption{Optical and infrared colours for the ATLAS9-ODFNEW from Castelli \& 
Kurucz (2003)
compared to empirical colours for G and K dwarfs. The colours are shown as a
function of ($V-K_{S}$) for [M/H] equal to +0.5 (black line), +0.0 (red line),
$-$0.5 (cyan line), $-$1.0 (green line), $-$1.5 (blue line), $-$2.0 (yellow
line). Points correspond to observed colours for the sample stars in the range
[M/H] $> 0.25$ (black), $-0.25 <$ [M/H] $\leq$ 0.25 (red), $-0.75 <$ [M/H]
$\leq -0.25$ (cyan), $-1.25 <$ [M/H] $\leq -0.75$ (green), $-1.75 <$ [M/H]
$\leq -1.25$ (blue). The metallicities given for the model are solar-scaled,
whereas [M/H] for the stars has been computed using eq. (\ref{overme}).  A
typical error bar for the points is shown in the upper left of each plot. The
model and empirical colours are generally in very good agreement.}
\label{KurCol}
\end{center}
\end{figure*}

\begin{figure*}
\begin{center}
\includegraphics[scale=0.86]{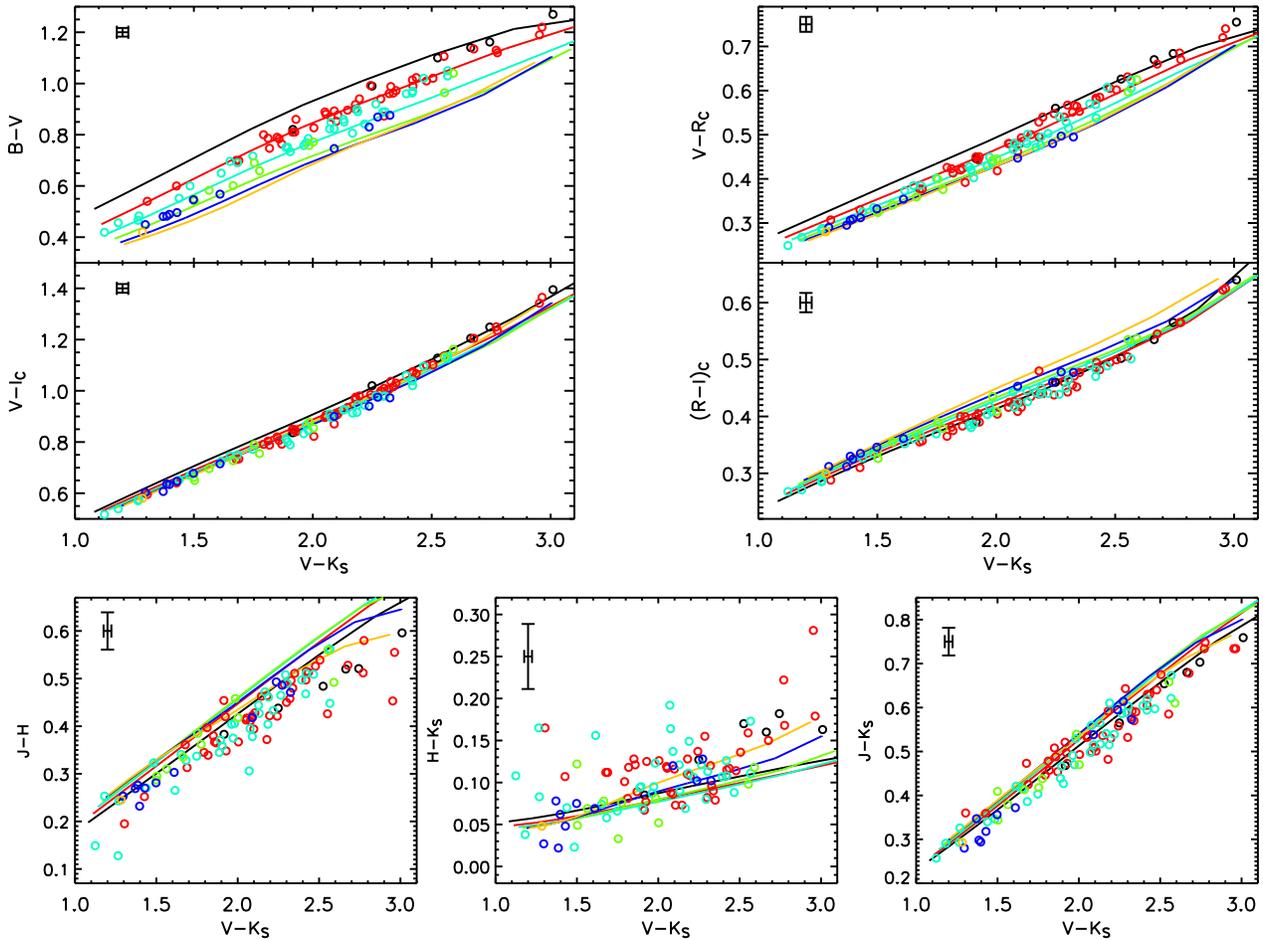}
\caption{As Figure \ref{KurCol} but for MARCS models, although the metallicity
range is different, since MARCS model are `$\alpha$-enhanced'. Lines of
constant [M/H] correspond to +0.5 (black line), +0.0 (red line), $-$0.35 (cyan
line), $-$0.69 (green line), $-$1.19 (blue line), $-$1.69 (yellow line). Points
correspond to observed colours for the sample stars in the range [M/H] $> 0.25$
(black), $-0.175 <$ [M/H] $\leq$ 0.25 (red), $-0.52 <$ [M/H] $\leq -0.175$
(cyan), $-0.94 <$ [M/H] $\leq -0.52$ (green), $-1.44 <$ [M/H] $\leq -0.94$
(blue) and [M/H] $\leq -1.44$ (yellow). Notice that MARCS model are
$\alpha$--enhanced for metallicities below the solar and [M/H] for stars and
models have been computed using eq. (\ref{overme}).}
\label{MARCSCol}
\end{center}
\end{figure*}

\begin{figure*}
\begin{center}
\includegraphics[scale=0.860]{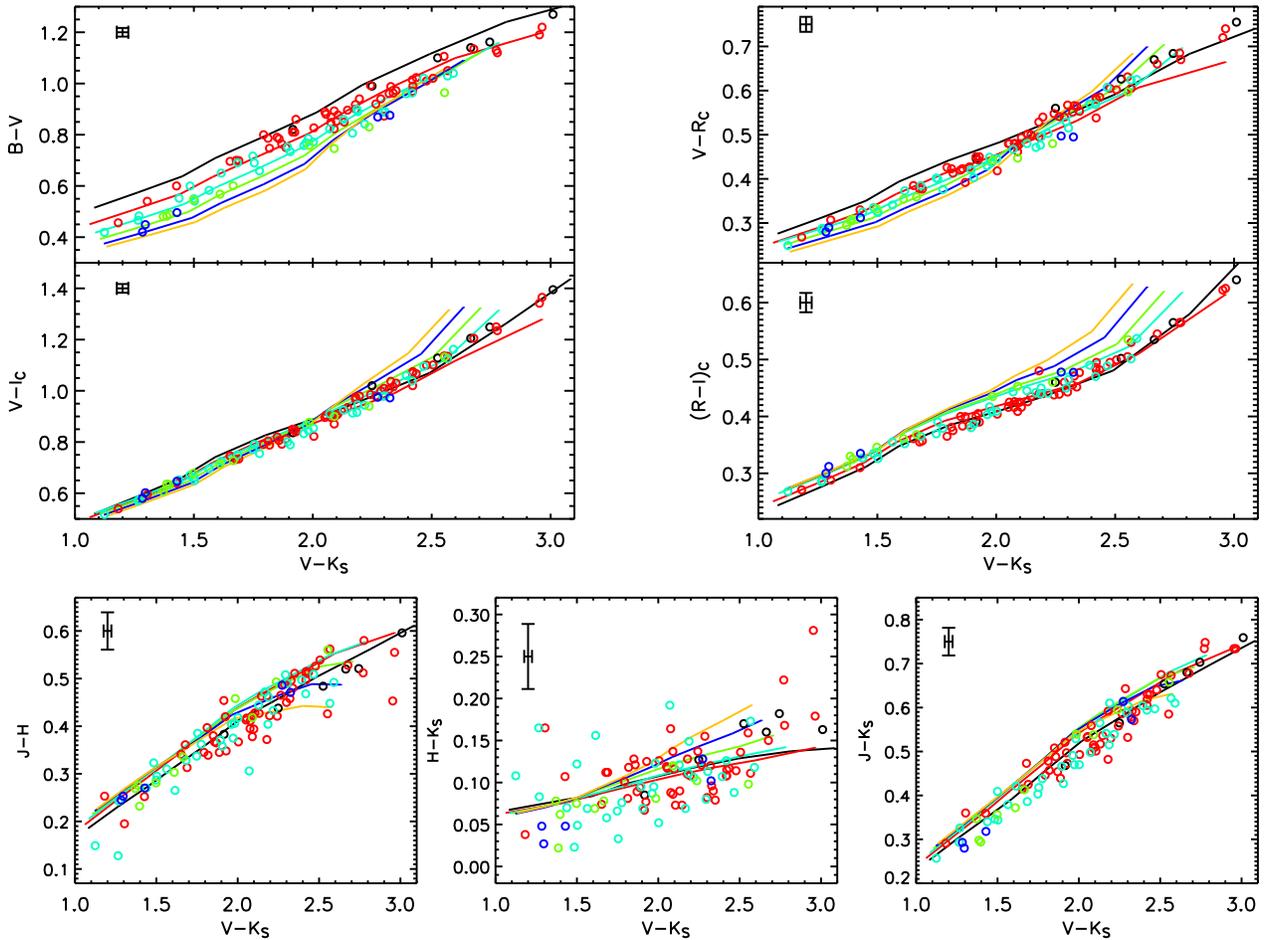}
\caption{Comparison of empirical and theoretical colours for G+K
dwarfs. Same schema as Figure \ref{KurCol} but for the BaSel 3.1 models.  The
BaSel 2.1 models compare similarly to the empirical data, but are not shown for
the sake of brevity (contact the author if interested).}
\label{WesCol}
\end{center}
\end{figure*}

\begin{figure*}
\begin{center}
\includegraphics[scale=0.7]{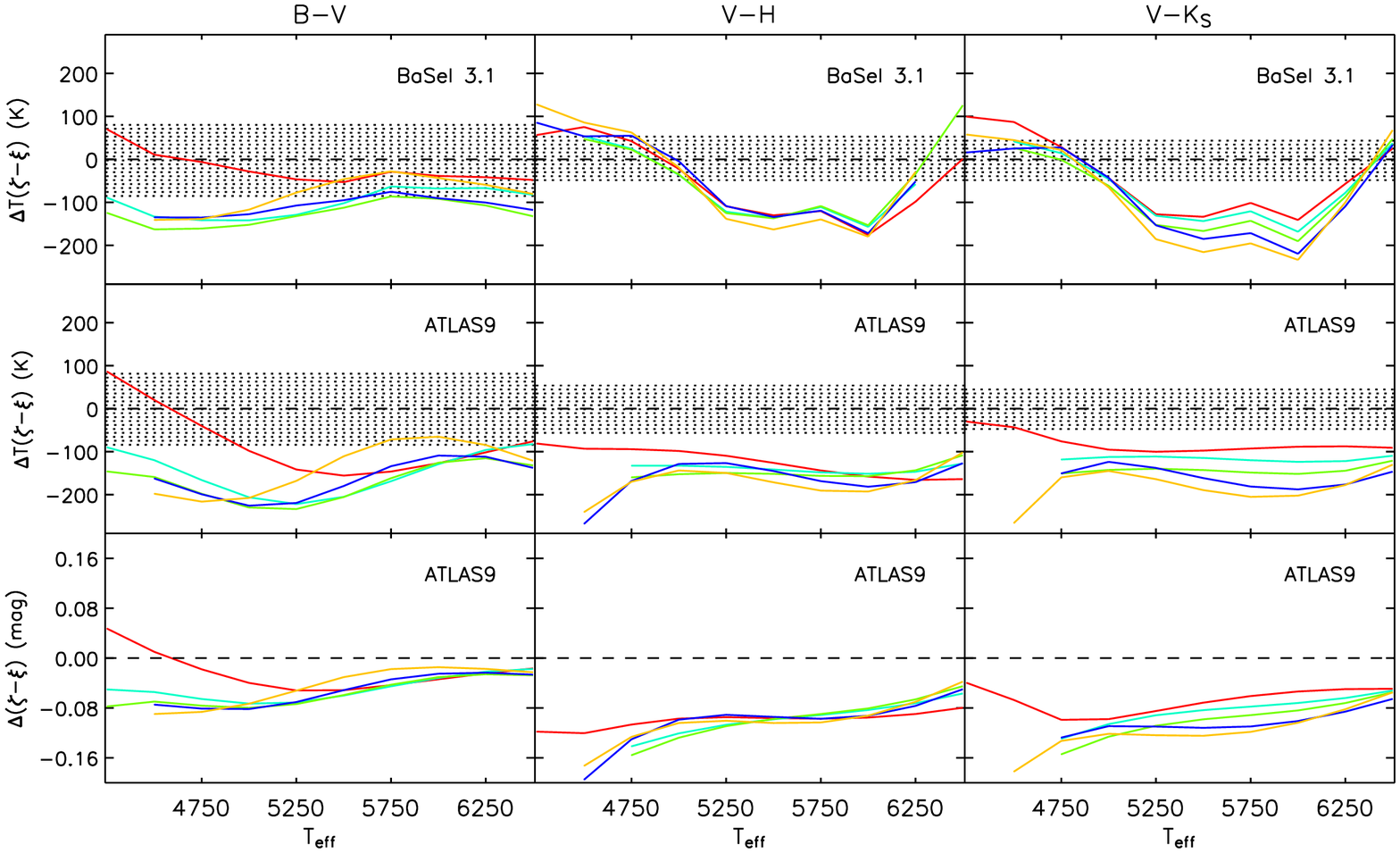}
\caption{{Effective temperatures recovered by means of the Ram\'irez \& 
    Mel\'endez
    empirical calibration are compared with the effective temperature of the
    corresponding BaSel 3.1 (first row) and ATLAS9 (second row) models used to
    generate synthetic photometry. $\zeta-\xi$ indicate the colours $B-V$,
    $V-H$ and $V-K_S$. The notation $\Delta \textrm{T}(\zeta-\xi)$ means
    temperature found by means of the $(\zeta-\xi)$ calibration minus the
    effective temperature of the model atmosphere used to generate the
    synthetic $\zeta$ and $\xi$ colours. The following metallicities are
    plotted: +0.0 (red), $-$0.5 (cyan), $-$1.0 (green), $-$1.5
    (blue) and $-$2.0 (yellow). The dashed area is the standard deviation of
    the empirical calibration in the corresponding colours. The third row shows
    the shift in colours required to set the ATLAS9 models on the empirical
    relation.}}
\label{modRAM}
\end{center}
\end{figure*}

\subsection{Reddening corrections}
\label{redden}

Interstellar absorption and reddening must be taken into account for a correct
derivation of stellar parameters, but these effects are negligible for our
sample stars, as we discuss in this section.

The distance distribution of the sample peaks around $\sim 30$ pc, the most
distant object being located at a distance of $\sim 70$ pc. For distances
closer than $50$ pc the polarimetric approach is extremely sensitive and can be
used as a lower limit even to the expectedly small amounts of dust (at least
for anisotropic particles).  Tinbergen (1982) and Leroy (1993b) have confirmed
the complete depletion of dust within $\sim 40$ pc from the Sun.  Using the
catalogue of Leroy (1993a) we have found polarimetric measurements for 21 stars
out of the 129 single and non-variable stars selected in Section \ref{double}.
The mean percentage polarization is $31 \times 10^{-5}$ which using the
Serkowski, Mathewson \& Ford (1975) conversion factor corresponds to a $E(B-V)$
of 0.0034 mag.

The Str\"omgen H$\beta$ index offers an alternative method for assessing the
reddening of individual stars in our sample, especially for the most distant
ones; 59 stars in our \emph{bona fide} single and non-variable sample have
H$\beta$ measurements (Hauck \& Mermilliod 1998) from which $E(b-y)$ was
derived using the intrinsic-colour calibration of Schuster \& Nissen (1989)
plus a small offset correction as noted by Nissen (1994). The reddening
distribution for these 59 stars peaks around $E(b-y)$ = 0.004, retaining the
same mean value (and never exceeding 0.015 for a single star) when we restrict
the sample to distances further than $50$ pc. $E(B-V)$ can then be calculated
using the standard extinction law or the relation $E(B-V) = 1.35 E(b-y)$
derived by Crawford (1975). Once the reddening $E(B-V)$ is known, the
extinction at any given wavelength can be determined using the standard
extinction law.  Given that the standard error in the Schuster \& Nissen
calibration is on the order of 0.01 in $E(b-y)$ and observational errors in
$uvby$ measurements are possible, the implied reddening corrections are below
the noise level. In addition to this there are indications (Knude 1979; Vergely
et al. 1998) that interstellar reddening is primarily caused by small dust
clouds causing it to vary in steps of 0.01--0.03 mag which confirm that the
small corrections found are consistent with the assumption of zero reddening
for the sample as a whole.

\section{Empirical colours versus model colours}
\label{Compa}

Theoretical stellar models predict relations between physical quantities such
as effective temperature, luminosity and stellar radius. From the observational
point of view these quantities are, in general, not directly measurable and
must be deduced from observed broad-band colours and magnitudes. In this
context, empirical colour-temperature relations (Bessell 1979; Ridgway et
al. 1980; Saxner \& Hammarback 1985; Di Benedetto \& Rabbia 1987; Blackwell \&
Lynas-Gray 1994, 1998; Alonso et al. 1996a; Ram\'irez \& Mel\'endez 2005a) and
colour-bolometric correction relations (Bessell \& Wood
1984; Malagnini et al. 1986; Bell \& Gustafsson 1989; Blackwell \& Petford
1991a; Alonso et al. 1995) are normally used. Such calibrations were usually
restricted to a specific type and/or population of stars, the only exception
being Alonso et al. (1995, 1996a) who give empirical relations for F0-K5 dwarfs
with both solar and sub-solar metallicity. Recently Ram\'irez \& Mel\'endez 
(2005a) have improved and extended the Alonso et al. (1996a)
colour-temperature-metallicity relations.

In any case, even empirical calibrations make use of model atmospheres to some
extent, though the model dependence is expected to be small (see Section
\ref{onmodel}). For example, the work of Alonso et al. (1996b) was based on
Kurucz' (1993) spectra, whereas the recent Ram\'irez \& Mel\'endez (2005b)
calibration is based on the original Kurucz spectra (i.e. without empirical
modifications) taken from Lejeune et al. (1997).

In this section we compare a number of stellar models (ATLAS9, MARCS, BaSel 3.1
and BaSel 2.1.) to our empirical data, mainly in two colour planes. In general,
colour-colour plots of the stars against all the model sets look quite
satisfactory. 

\subsection{ATLAS9, MARCS and BaSeL spectral libraries}
\label{libraries}

We have tested synthetic spectra from the ATLAS9-ODFNEW (Castelli \& Kurucz
2003), MARCS (Gustafsson et al. 2002), BaSel 2.1 and 3.1 (Lejeune et al. 1997
and Westera et al. 2002, respectively) libraries. All grids are given in steps
of $250$ K. Notice that, since we are working with dwarf stars, we assume
$\log(g)=4.5$ throughout. This assumption is reasonable given that
determinations of $\log (g)$ have a typical uncertainty of 0.2 dex or more
either by requiring FeI and FeII lines to give the same iron abundance or by
using Hipparcos parallaxes (eg. Bai et al. 2004).  This uncertainty covers the
expected range in $\log(g)$ on the main sequence (see Figure \ref{logg}). In
any case, a change of $\pm 0.5$ dex in the surface gravity implies differences
that never exceed a few degrees in derived effective temperature, as will be
seen (Section \ref{Felog}).

\subsubsection{ATLAS9-ODFNEW models}

The ATLAS9-ODFNEW models calculated by Castelli \& Kurucz include improvements
in the input physics, the update of the solar abundances from Anders \&
Grevesse (1989) to Grevesse \& Sauval (1998) and the inclusion of new molecular
line lists for TiO and $\mathrm{H}_{2}$O.  The metallicities cover $-2.5 \le$
[M/H] $\le +0.5$ with solar-scaled abundance ratios. A microturbulent velocity
$\xi=2\;\mathrm{km}\;\mathrm{s}^{-1}$ and a mixing length parameter of 1.25 are
adopted. The extension of these models to include also $\alpha$-enhanced
chemical mixtures is ongoing.  However, we do not expect large differences
among different models as a result of $\alpha$-enhancement, as explained in
Section \ref{onmodel}. In the remainder of the paper we refer to the
ATLAS9-ODFNEW model simply as ATLAS9.

\subsubsection{MARCS models}

The new generation of MARCS models (Gustafsson et al. 1975; Plez et al. 1992)
now includes a much-improved treatment of molecular opacity (Plez 2003;
Gustafsson et al. 2002), chemical equilibrium for all neutral and singly
ionised species as well as some doubly ionised species, along with about 600
molecular species. The atomic lines are based on VALD (Kupka et al. 1999). The
models span the metallicity range $-5.0 \leq \textrm{[Fe/H]} \leq +1.0 $. We
have used models with solar relative abundances for [Fe/H] $\geq 0$ and
enhanced $\alpha$-elements abundances for [Fe/H] $< 0$. The fraction of
$\alpha$--elements is given by the model and [M/H] has been computed using eq.
(\ref{overme}).

\subsubsection{The BaSel libraries}

The BaSel libraries are based on a posteriori empirical corrections to hybrid
libraries of spectra, so as to reduce the errors of the derived synthetic
photometry.  BaSel 2.1 (Lejeune et al. 1997) is calibrated using solar
metallicity data only and it is known to be less accurate at low metallicities
([Fe/H] $<1$) especially in the ultraviolet and infrared (Westera et
al. 2002). The
colour calibration has recently been extended to non-solar metallicities by
Westera et al. (2002). Surprisingly, they found that a library that matches
empirical colour-temperature relations does not reproduce Galactic
globular-cluster colour--magnitude diagrams, as a result of which they propose
two different versions of the library. In what follows we have used the library
built to match the empirical colour-temperature relations.  For both BaSel
libraries, it is obvious that the transformations used do not correct the
physical cause of the discrepancies.  However, it is worthwhile to check
whether they do give better agreement with the empirical relations.

All libraries have been tested in the range $4250 \le T_{\mathrm eff} \le 6500$
K and $-2.0 \le$ [M/H] $\le +0.5$, the intervals covered by our sample of
stars.

\subsection{Comparison of model and empirical colours}

\begin{figure*}
\begin{center}
\includegraphics[scale=0.7]{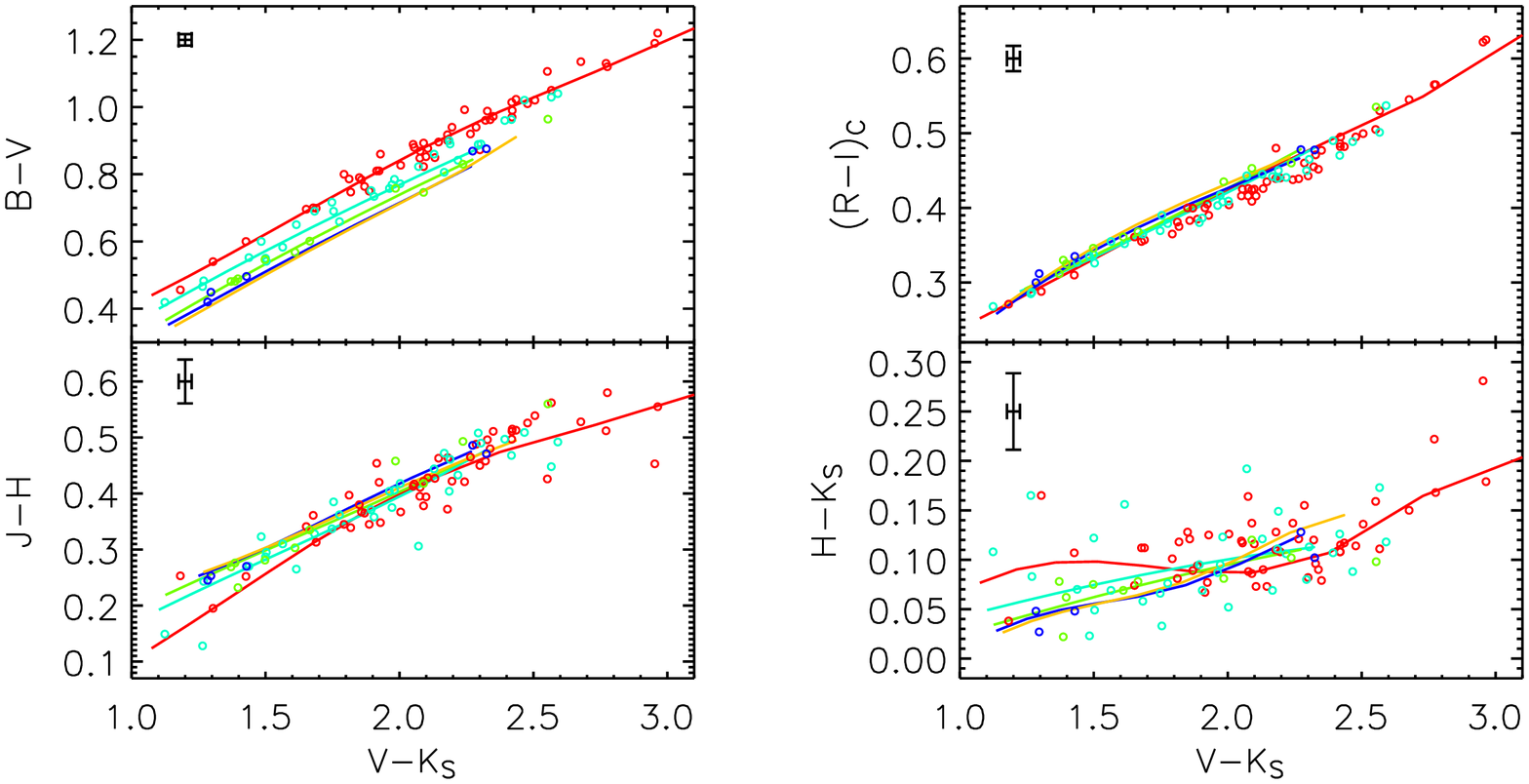}
\caption{{Lines represent the colour-colour relations derived from the
    empirical colour-temperature calibrations of Ram\'irez \& 
    Mel\'endez (2005a). Points correspond to our sample of stars. Colours and
    metallicity ranges for lines and points are same as in Figure 
    \ref{KurCol} except that metallicities of +0.5 have been excluded since
    the empirical relations do not hold in that range. In the optical the
    empirical relations fit the observed colours with an accuracy comparable to
    the theoretical ones (compare with Figure \ref{KurCol}). However in the
    infrared, for metallicities around the solar, empirical relations 
    oscillate in reproducing the observed trend. 
    This is especially interesting in light of the puzzling infrared colours of 
    the Sun and the shifts required to cool down the temperature scale as 
    discussed in Section \ref{checking}.}}
\label{empobs}
\end{center}
\end{figure*}
All our colour-colour comparisons are as a function of $(V-K_{S})$. This
approach highlights the behaviour of colours as function of temperature and
also make more straightforward the comparison with the empirical
colour-temperature relations in the Section \ref{emp}. 
$(V-K_{S})$ has a large range
compared to its observational uncertainty and it is known to be metallicity
insensitive (e.g. Bell \& Gustafsson 1989 and Figure \ref{colteplot}).

All models are found to be in good overall agreement with the observations. In
particular, the improvement in the input physics of the latest ATLAS9 and 
MARCS models show excellent agreement with the optical data (Figures 
\ref{KurCol} and \ref{MARCSCol}) and as good as the semi-empirical libraries 
in the other bands, so that we do not find any specific reason to prefer the 
use of semi-empirical libraries. 

We expect the IR colours to be very important both in observational terms and
in theoretical modelling.  Recently, Fr\'emaux et al. (2006) have presented a
detailed comparison of theoretical spectra (from the NeMo library) in $H$ band
with observed stellar spectra at resolution around 3000. They show that in the
infrared range, although the overall shape of the observed flux distribution is
matched reasonably well, individual features are reproduced by the theoretical
spectra only for stars earlier than mid F type. For later spectral types the
differences increase and theoretical spectra of K type stars have
systematically weaker line features than those found in observations. They
conclude that these discrepancies stem from incomplete data on neutral atomic
(and to a minor extent to molecular) lines.  

The synthetic optical colours agree well with observed ones and models also
predict the correct metallicity dependence. In the infrared, models fit
satisfactorily the overall trend given by the observed colours, however the 
scatter in the data prevent firm conclusions. In $J-H$ and $J-K_S$ the model
colours seem 
to be slightly offset by $\sim 0.05$ mag, whereas in $H-K_S$ the disagreement 
is somewhat smaller though it depends on the models considered and it seems to
increase going to cooler temperatures.

\subsection{Empirical colour-temperature relations compared to observed and 
            synthetic photometry}\label{emp}

We now compare empirical colour-temperature calibrations with synthetic and
observed photometry. To this purpose, we use the calibration recently proposed 
by Ram\'irez \& Mel\'endez (2005a).

In Fig. \ref{modRAM} we plot the difference between the temperature of the
model atmospheres (BaSel 3.1 and ATLAS9) and the temperature expected for their
colours according to the empirical Ram\'irez \& Mel\'endez relations.
Interestingly, all models fail to set onto the empirical temperature scale
being at least 100 K hotter in optical and infrared colours. The disagreement
is particularly strong at low metallicity as noticed by Ram\'irez \&
Mel\'endez.  The semi-empirical libraries of Lejeune et al. (1997) and Westera
et al.  (2002), though in principle optimized to reproduce colour-temperature
relations, provide only a marginal improvement with respect to the latest
purely theoretical ATLAS9 and MARCS libraries. The temperature offset can be
translated into a shift in colours needed to set the models onto the empirical
scale. This is done for the ATLAS9 model in the last row of Figure
\ref{modRAM}; in terms of colours, the mismatch corresponds to $\sim 0.1$ mag.

Empirical calibrations depend on metallicities and colour indices only.  The
adopted absolute calibration can not be the main cause of the disagreement
since all colours (in the optical) scale accordingly. In the optical the
adopted Vega model that sets the absolute calibration has been accurately
tested by Bohlin \& Gilliland (2004) (see Appendix A). Although we adopt a
different absolute calibration in the optical and in the infrared (see Appendix
A), this could at most affect the colour-metallicity-temperature relations in
$V-J$, $V-H$ and $V-K_S$. Even using in the infrared the same absolutely
calibrated Vega model adopted in the optical does not eliminate the
disagreement and offsets of 100~K still persist.

We have also verified that the choice of zero-points plays a negligible role:
differences in the optical zero-points based on Vega or on the average of Vega
and Sirius (see Appendix A) lead to mean temperature differences of 10 to 30
K. Besides being much smaller than the uncertainty in the colour-temperature
relations, such differences go in the direction of worsening the disagreement.

As to the IR zero-points, we checked their effects by generating IR synthetic
magnitudes in the TCS and Johnson system (where the Vega zero-points are
different from those deduced from 2MASS photometry) and compared the model
predictions to the Alonso et al. (1996a) empirical calibration in these bands;
typical discrepancies of more than 100~K still persist.

The causes for the discrepancy must therefore be model deficiencies and/or
inaccuracies in the adopted empirical relations.

In fact we caution that also empirical relations may hide some inadequacies.
For instance, the Ram\'irez \& Mel\'endez calibration predicts slightly bluer
colours for the Sun than the recent completely empirical determination of
Holmberg, Flynn \& Portinari (2006). We can also combine and invert the
Ram\'irez \& Mel\'endez' colour-temperature calibrations, to derive the
corresponding \emph{empirical} colour-colour relations and compare them to the
\emph{observed} colours for our sample of stars. Interestingly the agreement is
not always good: the solar metallicity is slightly offset with respect to the
data in $(R-I)_C$ band and in the infrared tends to oscillate (see Figure
\ref{empobs}). This underlines that there is still room for improvement also in
the empirical relations.

\section{An implementation of the IRFM}
\label{irfm}

In this section we have used $BV(RI)_C JHK_S$ photometry as the basic
observational information to derive bolometric fluxes and effective
temperatures of our 104 G and K dwarfs. Our approach follows the InfraRed Flux
Method (IRFM) used in the extensive work of Alonso, Arribas \& Mart\'inez-Roger
(1995, 1996b). Ours is a new independent implementation of the IRFM, applied to
our 104 G and K type dwarfs.

Our implementation differs from Alonso et al (1995, 1996), as we base its
absolute calibration on a synthetic spectrum of Vega (as described in detail in
Appendix A) rather than on semi-empirical measurements (Alonso, Arribas \&
Mart\'inez-Roger, 1994). The use of absolutely calibrated synthetic spectra
rather than ground-based measurements traces back to Blackwell et al. (1990,
1991a), who concluded that atmospheric models of Vega offered higher precision
than observationally determined absolute calibrations. This approach has been
adopted in the extensive work of Cohen and collaborators and also Bessell,
Castelli \& Plez (1998), who concluded that model atmospheres are more reliable
than the near-infrared absolute calibration measurements. Models are nowadays
sophisticated enough to warrant detailed comparisons between observation and
theory and the latest space based measurements confirm the validity of the
adopted absolute calibration (see Appendix A).  We also differ from the recent
work of Ram\'irez \& Mel\'endez (2005b) in using the most recent model
atmospheres and absolute calibration available.  Furthermore we use only direct
observational data, explicitly avoiding the use of colour calibrations from
previous studies to infer the bolometric luminosity.

Following Alonso et al. (1995), the flux outside the wavelength range covered
by our photometry has been estimated using model atmospheres. Since the
percentage of $F_{bol}$ measured in the directly observed bands range from
$\sim 70$\% to $\sim 85$\% depending on the star, the dependence of the
estimated bolometric flux on the model is small.

In Section \ref{bolome} and \ref{irfmSec} we give a detailed description of our
implementation of the IRFM, mainly following the formulation and terminology
adopted by Alonso et al. (1995, 1996b). 

Our procedure can be summarized as follows. Given the metallicity, the surface
gravity (assuming $\log(g)=4.5$ for all our dwarfs, see Section
\ref{libraries}) and an initial estimate for the temperature of a star, we have
interpolated over the grid of model atmospheres to find the spectrum that best
matches these parameters. This spectrum is used to estimate that fraction of
the bolometric flux outside our filters $(BV(RI)_CJHK_S)$, i.e. the `bolometric
correction'. The bolometric flux is determined from the observations, including
the bolometric correction. A new effective temperature $T_{\mathrm eff}$ can be
computed by means of the IRFM. This temperature is used for a second
interpolation over the grid, and the procedure is iterated until the
temperature converges to within 1 K (typically within 10 iterations).

We have tested the results using ATLAS9, MARCS, BaSel 2.1 and 3.1 libraries. A
detailed discussion on the the dependence on the adopted model is given in
Section \ref{onmodel}.

In the following sections we describe our implementation of the IRFM effective
temperature scale.

\subsection{Bolometric fluxes}
\label{bolome}

As mentioned above, we use grids of synthetic spectra to bootstrap our IRFM.  
The grids of synthetic spectra all have a resolution of 250 K in
temperature whereas the resolution in metallicity depends on the library, but
with typical steps of 0.25 or 0.50 dex.

For any given star of overall metallicity [M/H] (c.f. Eq. (\ref{overme}), we
interpolate over the grid of synthetic spectra in the following way. First, we
use the $T_{\mathrm eff}:(V-K)$ calibration of Alonso et al. (1996a) to obtain
an initial estimate of the effective temperature $T_{\mathrm eff,0}$ of the
star. Then we linearly interpolate in $\log (T_{\mathrm eff})$ bracketing our
temperature estimate $T_{\mathrm eff,0}$ at two fixed values of metallicity,
which bracket the measured [M/H] of the star.  A third linear interpolation is
finally done in metallicity in order to obtain the desired synthetic spectrum.

Having obtained the spectrum $F(\lambda)$ that better matches the physical
parameters of the star, for any given band $\zeta$ (running from $B$ to
$K_{S}$), we convolve it through the transmission curve $T_{\zeta}(\lambda)$ of
the filter and associate the resulting $\mathcal{F}_{\zeta}(\textrm{model})$
with its effective wavelength $\lambda_{\mathrm eff}$ (see Appendix B).

We then compute the flux covered by the passbands $B$ to $K_{S}$ for the model
star ($\mathcal{F}_{B-K_S} (\textrm{model})$).  When the latter is compared
with the bolometric flux for the same model star
($\mathcal{F}_{Bol}(\textrm{model})$) we can obtain an estimate of the fraction
of the flux $\mathcal{C}$ encompassed by our filters:

\begin{equation}
\label{cfact}
\mathcal{C}=\frac{\mathcal{F}_{B-K_S}(\textrm{model})}{\mathcal{F}_{Bol}
(\textrm{model})}.
\end{equation}

We now have the correction factor $1/\mathcal{C}$ for the missing flux of a
given star, we use its observed $BV(RI)_C JHK_S$ magnitudes ($m_{\zeta}$) to
calculate the flux as it arrives on the Earth:

\begin{equation}
\label{earthflx}
\mathcal{F}_{\zeta}(\textrm{Earth})=\mathcal{F}_{\zeta}^{std}(\textrm{Earth})
10^{-0.4(m_{\zeta}-m_{\zeta}^{std})},
\end{equation}

where $\mathcal{F}_{\zeta}^{std}(\textrm{Earth})$ is the absolute calibrated
flux on the Earth of the standard star and $m_{\zeta}^{std}$ is its observed
magnitude. The observed magnitudes and the absolute calibration of the standard
star are those given in Table \ref{vzpo} and \ref{absflux} respectively and
play a key role in determining both bolometric flux and effective temperature
as we discuss in Section \ref{AbsCalSys}.

The flux at the Earth for each band is once again associated with the
corresponding effective wavelength of the star, and a simple integration leads
to $\mathcal{F}_{B-K_S}(\textrm{Earth})$. The latter is then divided by the
$\mathcal{C}$ factor as defined in eq. (\ref{cfact}) in order to obtain the
bolometric flux measured on the Earth $\mathcal{F}_{Bol}(\textrm{Earth})$.

\subsection{Effective Temperatures}
\label{irfmSec}

The effective temperature $T_{\mathrm eff}$ of a star satisfies the
Stefan-Boltzman law $\mathcal{F}_{Bol}=\sigma T_{\mathrm eff}^{4}$, where
$\mathcal{F}_{Bol}$ is the bolometric flux on the surface of the star. Only the
bolometric flux on the Earth is measurable, and we must take into account the
angular diameter ($\theta$) of the star. Thus:

\begin{equation}\label{bolflux}
\mathcal{F}_{Bol}(\textrm{Earth}) = \left(\frac{\theta}{2}\right)^2 \sigma 
T_{\mathrm eff}^{4}. 
\end{equation}

The way to break the degeneracy between effective temperature and angular
diameter in the previous equation is provided by the InfraRed Flux Method
(Blackwell \& Shallis 1977; Blackwell, Shallis \& Selby 1979; Blackwell,
Petford \& Shallis 1980).  The underlying idea relies on the fact that whereas
the bolometric flux depends on both angular diameter
and effective temperature (to the fourth power), the monochromatic flux
\emph{in the infrared} at Earth --
$\mathcal{F}_{\lambda_{\textrm{\tiny{IR}}}}(\textrm{Earth})$ -- depends on the
angular diameter but only weakly (roughly to the first power) on the effective
temperature:

\begin{equation}
\label{monflux}
\mathcal{F}_{\lambda_{\textrm{\tiny{IR}}}}(\textrm{Earth}) = 
\left(\frac{\theta}{2}\right)^2 \phi(T_{\mathrm eff},g,\lambda_{\textrm{\tiny{IR}}}), 
\end{equation}

where $\phi(T_{\mathrm eff},g,\lambda_{\textrm{\tiny{IR}}})$ is the monochromatic
\emph{surface} flux of the star. The ratio $\mathcal{F}_{Bol}(\textrm{Earth})/
\mathcal{F}_{\lambda_{\textrm{\tiny{IR}}}}(\textrm{Earth})$ defines what is
known as the observational $R$-factor ($R_{obs}$), where the dependence on
$\theta$ is eliminated. In this sense the IRFM can be regarded as an extreme
example of a colour method for determining the temperature.  By means of
synthetic spectra it is possible to define a theoretical counterpart
($R_{theo}$) on the \emph{surface} of the star, obtained as the quotient
between the integrated flux ($\sigma T_{\mathrm eff}^4$) and the monochromatic 
flux in
the infrared $\mathcal{F}_{\lambda_{\textrm{\tiny{IR}}}}(\textrm{model})$.

The basic equation of the IRFM thus reads:

\begin{equation}
\label{irfmeq}
\frac{\mathcal{F}_{Bol}(\textrm{Earth})}
{\mathcal{F}_{\lambda_{\textrm{\tiny{IR}}}}(\textrm{Earth})} = \frac{\sigma
T_{\mathrm eff}^4}{\mathcal{F}_{\lambda_{\textrm{\tiny{IR}}}}(\textrm{model})}
\end{equation}

and can be immediately rearranged to give the effective temperature,
eliminating the dependence on $\theta$.

The monochromatic flux $\mathcal{F}_{\lambda_{\textrm{\tiny{IR}}}}
(\textrm{Earth})$ is obtained from the infrared photometry using the following
relation:

\begin{equation}
\label{qmono}
\mathcal{F}_{\lambda_{\textrm{\tiny{IR}}}}(\textrm{Earth})=
q(\lambda_{\textrm{\tiny{IR}}})\;\mathcal{F}_{\lambda_{\textrm{\tiny{IR}}}}
^{std}(\textrm{Earth})\;10^{-0.4(m_{\textrm{\tiny{IR}}}^{}-m_{\textrm{\tiny{IR}}}^{std})},
\end{equation}

where $q(\lambda_{\textrm{\tiny{IR}}})$ is a correction factor to determine
monochromatic flux from broad-band photometry, $m_{\textrm{\tiny{IR}}}$ is the
infrared magnitude of the target star, $m_{\textrm{\tiny{IR}}}^{std}$ and
$\mathcal{F}_{\lambda_{\textrm{\tiny{IR}}}}^{std}(\textrm{Earth})$ are the
magnitude and the absolute monochromatic flux of the standard star in the same
infrared band (see Table \ref{vzpo} and \ref{absflux}). For the standard and
the target star $\lambda_{IR}$ refers to their respective effective wavelength
in the given infrared band, as we discuss in more detail, together with the
definition of the $q$--factor, in Appendix B.

The IRFM is often applied using more than one infrared wavelength and different
$T_{\mathrm eff}$ are obtained for each wavelength. Ideally, the derived 
temperatures
should of course be independent of the monochromatic wavelength used.  In our
case we have used $J, H, K_S$ photometry and the corresponding effective
wavelengths.  

\begin{figure}
\begin{center}
\includegraphics[scale=0.21]{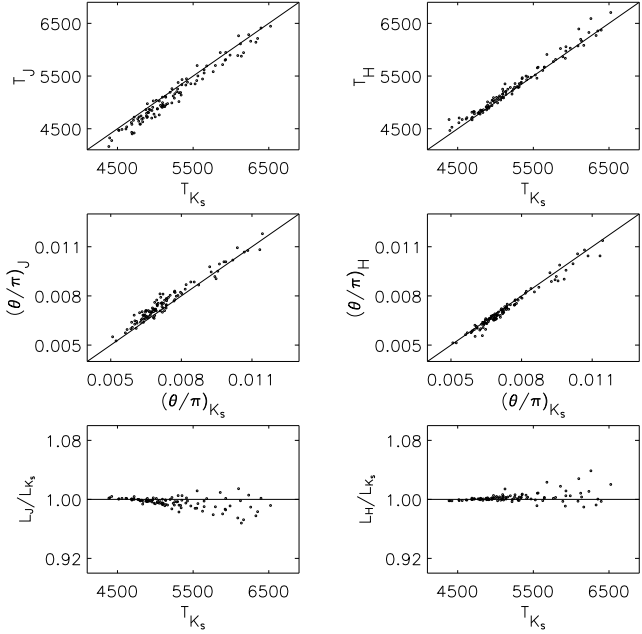}
\caption{Comparison of effective temperatures, angular diameters and bolometric
luminosities obtained in the three infrared bands used. Angular diameters have 
been scaled using the stellar parallaxes from Hipparcos, although the 
parallaxes do not affect the comparison since all bands scale 
accordingly. The luminosities of the stars are in good agreement
across all the IR bands (because luminosity is also tied down by the optical
data as discussed in the text). Offsets of this size are likely to stem from 
the difficulties in setting the zero points of the absolute calibration. 
Horizontal and diagonal lines with slope 1 are intended to guide the eye.}
\label{jhk}
\end{center}
\end{figure}

We have tested the systematic differences in temperatures, luminosities and
diameters obtained when \emph{only one infrared band at a time} is used for the
convergence of the IRFM. The results are shown in Figure \ref{jhk}.

The $J$ band returns temperatures that are systematically cooler than those
found from $K_{S}$, which translates into greater angular diameters.  The $H$
band, returns temperatures that are systematically hotter and therefore smaller
angular diameters than in $K_{S}$.  These systematic differences do not appear
to depend on the temperature range, at least between 4500--6500~K.

The systematic offsets could be traced back to the absolute calibration in
different bands. We further point out that for the sake of this test, the
convergence on $T_{\mathrm eff}$ in the IRFM is required in one band only and
in Figure \ref{jhk} the differences are thus exaggerated.  The discrepancy can
reach a few percent in temperatures and diameters, whereas for bolometric
luminosities is usually within 2\% though it increases with increasing
temperature (Figure \ref{jhk}). This is due to the fact that the derived
bolometric luminosity is constrained from the full optical and infrared
photometry and absolute calibration, whereas for temperature and angular
diameter, when relying on one IR band only, the absolute calibration in that
band strongly affects the results.

Alonso et al. (1996b) concluded that the consistency of the three infrared 
bands they used ($J,H,K$) was good over 4000 K, but decided to use only $H$ and
$K$ below 5000 K and only $K$ below 4000 K and basically the same is done by
Ram\'irez \& Mel\'endez (2005b).
We have decided to use the temperature returned in all three bands also below
5000~K, since systematic differences are not too large even below 5000~K, the
cooler temperatures in $J$ band being compensated by the hotter ones in $H$
band.

\begin{figure}
\begin{center}
\includegraphics[scale=0.22]{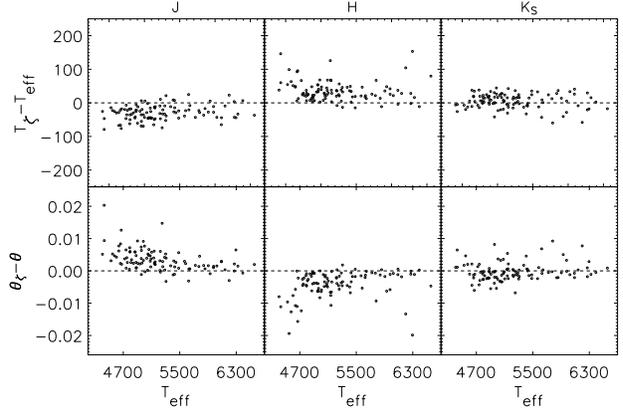}
\caption{{Differences as a function of temperature between the final adopted
          values and those obtained in each $\zeta=J,H,K_S$ band when the 
          convergence is done averaging the three IR values.}}
\label{jhk2}
\end{center}
\end{figure}

At each iteration the temperature used for the convergence is the average of
the three IR values weighted with the inverse of their errors ``j\_,'' ``h\_,''
and ``k\_msigcom''. This temperature is then used to select a new model
interpolating over the grid of synthetic spectra as described in Section
\ref{bolome} and a new bolometric flux and temperature are then computed. The
procedure is iterated until the average temperature converges within 1 K.  As
can be appreciated from Figure \ref{jhk2} the systematic
differences between the three bands are much reduced. Ideally the ratio of the
temperatures determined in the three bands should be unity. The three mean
ratios $T(K_S)/T(J)$, $T(K_S)/T(H)$ and $T(H)/T(J)$ together with their
standard deviations, are 1.0069 ($\sigma=0.77$\%), 0.9943 ($\sigma=0.82$\%) and
1.0126 ($\sigma=0.89$\%) and this also confirms the quality of the adopted
absolute calibration. The mean standard deviation for the three infrared
temperatures is 39~K, reflecting the different band sensitivity to the
temperature, the effect of the absolute calibration in different bands as well
as the observational errors in the infrared photometry.

\subsection{Internal accuracy and dependence on the adopted library}
\label{onmodel}

We have probed the accuracy and the internal consistency of the procedure
described in Sections \ref{bolome} and \ref{irfmSec} generating synthetic
magnitudes in the range $4250 \le T_{\mathrm eff} \le 6500$ and $-2.0 \le
\textrm{[Fe/H]} \le 0.5$ and testing how the adopted method recovers the
temperatures and luminosities of the underlying synthetic spectra.  The
accuracy is excellent given that in all three infrared bands the IRFM recovers
the right temperature within 1 K and the bolometric luminosity (i.e.\ the
theoretical value $\sigma T_{\mathrm eff}^4$) within 0.1\% and 0.04\% for the
ATLAS and MARCS models respectively, i.e.\ at the level of the numerical
accuracy of direct integration of the bolometric flux from the spectra.  This
confirms that the interpolation over the grid introduces no systematics and
even a poor initial estimate of the temperature does not affect the method.

We expect the dependence of the results on the adopted synthetic library to be
small, since most of the luminosity is actually observed photometrically and
the model dependence for the $q$-factor and $R_{theo}$ (see Section
\ref{irfmSec}) is weak because we are working in a region of the spectrum largely
dominated by the continuum. The method is in fact more sensitive to the adopted
absolute calibration that governs $R_{obs}$.

The principal contributor to continuous opacity in cool stars is due to
$\textrm{H}^{-}$. Blackwell, Lynas-Gray \& Petford (1991b) have shown that by
using more accurate opacities with respect to previous work, temperatures
increased by 1.3\% and angular diameters decreased up to 2.7\%, the effect
being greatest for cool stars. Even though the dependence of the results on the
adopted synthetic library is small, the use of absolute calibrations derived
using the most up-to-date model atmospheres (see Appendix A) is of primary
importance.  Though we use a great deal of observational information, the new
\emph{opacity distribution function} (ODF) in the adopted grid of model
atmospheres is particularly important. With respect to the more accurate but
computationally time consuming \emph{opacity sampling} (OS), older ODF models
can underestimate the IR flux by few percent, translating into cooler
effective temperatures (Grupp, 2004a). Likewise, the better opacities in the UV
lead to an increased flux in the visual and infrared region. This directly
affects the ratio between the bolometric and monochromatic flux used for the
Infrared Flux Method (see Section \ref{irfmSec}), so that the latest model
atmosphere have to be preferred (Megessier 1994; 1998).

\begin{figure}
\begin{center}
\includegraphics[scale=0.28]{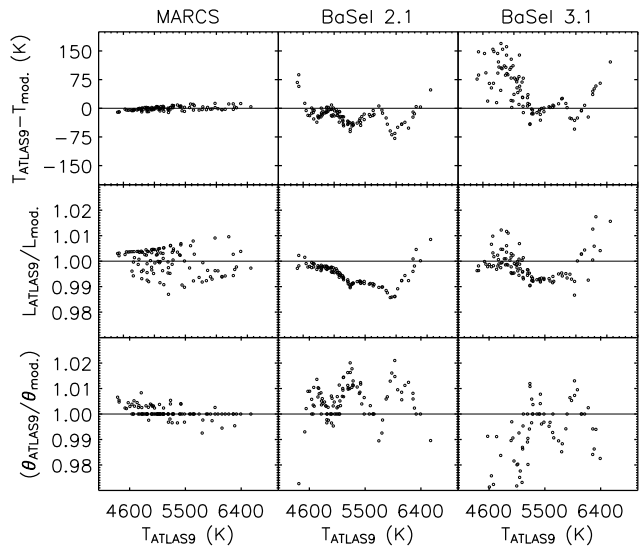}
\caption{Comparison between final effective temperatures, bolometric
    luminosities and angular diameters for our sample stars, when different
    synthetic libraries (with respect to ATLAS9) are used.}
\label{model}
\end{center}
\end{figure}

For all our 104 G--K dwarfs we have tested how the recovered parameters change
by using ATLAS9, MARCS, BaSel 2.1 and 3.1 libraries.  The comparison in
temperatures, luminosities and angular diameters is shown in Figure
\ref{model}.

The ATLAS9 and MARCS models show a remarkably good agreement through all the
temperature range, with differences in the temperature that never exceed a few
degrees (or 0.2\%) and those in the bolometric luminosities and the angular
diameters being always within 1\%. Though not a proof of the validity of the
models, it is very encouraging that the sophisticated physics implemented in
independent spectral synthesis codes shows such a high degree of consistency.

The BaSel models show bigger differences with respect to ATLAS9.  The BaSel 2.1
model tends to give slightly hotter
temperatures, higher luminosities and smaller angular diameters when compared
to ATLAS9, though the systematic is oscillating (reflecting the underlying
continuum adjustments of the semi-empirical BaSel libraries).

We conclude by noting that in the range of temperatures and luminosities
studied the dependence of our results on the adopted model never exceeds a few
percent and for the latest models (MARCS and ATLAS9) the agreement is well
within 1\%. Hence we will only present and discuss results based on ATLAS9 in
the following.

\subsection{Evaluation of the errors: rounding up the usual suspects}

Though the IRFM is known to be one of the most accurate ways of determining
stellar temperatures without direct angular diameter measurements, its
dependence on the observed photometry and metallicities, on the absolute
calibration adopted and on the library of synthetic spectra make the evaluation
of the errors not straightforward.  In this work we have proceeded from first
principles and made use of high accuracy observations only, thus facilitating
the estimation of errors and possible biases.  Since we made use of photometric
measurements in many bands, we expect the dependence on photometric errors to
be rather small, small random errors in different bands being likely to
compensate each other.  We have also shown in the previous section that when
the latest model atmospheres are adopted, the model dependence is below the one
percent level.  The effect of the absolute calibration can be regarded as a
systematic bias; once the absolute calibration in different bands is chosen,
all temperatures scale accordingly. As we show in Appendix A the adopted
absolute calibration has been thoroughly tested.

\subsubsection{Photometric observational error}
\label{MC}

The photometric errors in different bands as well as errors in metallicity are
likely to compensate each other. In order to check this, for each star we have
run 1000 MonteCarlo simulations, assigning each time random errors in
$BV(RI)_{C}JHK_S$ and [Fe/H]. The errors have been assigned with a normal
distribution around the observed value and a standard deviation of 0.01 mag in
$BV(RI)_{C}$ and of ``j\_,'' ``h\_,'' ``k\_msigcom'' in $JHK_{S}$. Errors in
[Fe/H] are those given in the papers from which we have collected abundances.

The standard deviation of the resulting temperatures obtained from the
MonteCarlo for each star is typically 11~K and never exceeds 19~K.  Bolometric
luminosities have a mean relative error of only 0.5\% and never exceeding
0.8\%; for angular diameters the mean error sets to 0.4\% and never exceeds
0.5\%.

\subsubsection{Metallicity and surface gravity dependence}
\label{Felog}

\begin{figure}
\begin{center}
\includegraphics[scale=0.22]{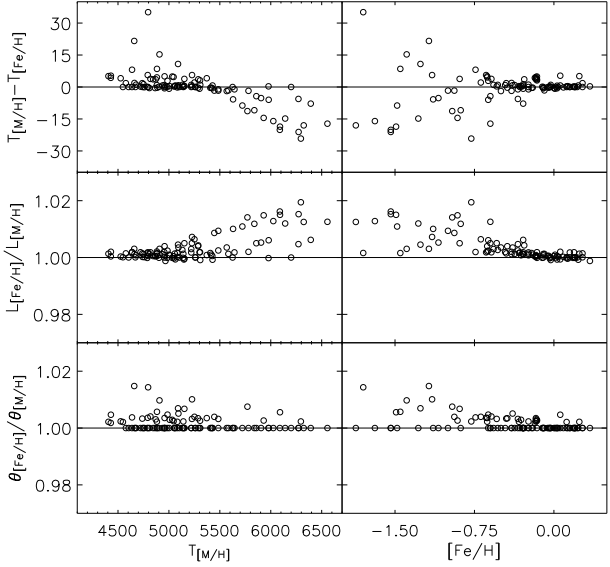}
\caption{{Differences between temperatures, bolometric luminosities and angular
    diameters when the [M/H] or [Fe/H] are used.}}
\label{FeMe}
\end{center}
\end{figure}

The interpolation within the grid of synthetic spectra is as a function of 
[M/H]. We have checked the effect of $\alpha$--elements by making the 
interpolation as a function of [Fe/H] instead, finding very little differences 
in the derived fundamental parameters around the solar values. 
At lower metallicities the effect of the $\alpha$--elements start to be 
important as expected (Figure \ref{FeMe}). The effect is in any case, always 
within 20~K in $T_{\mathrm eff}$ and within 2\% in bolometric luminosity and 
$\theta$.

\begin{figure}
\begin{center}
\includegraphics[scale=0.24]{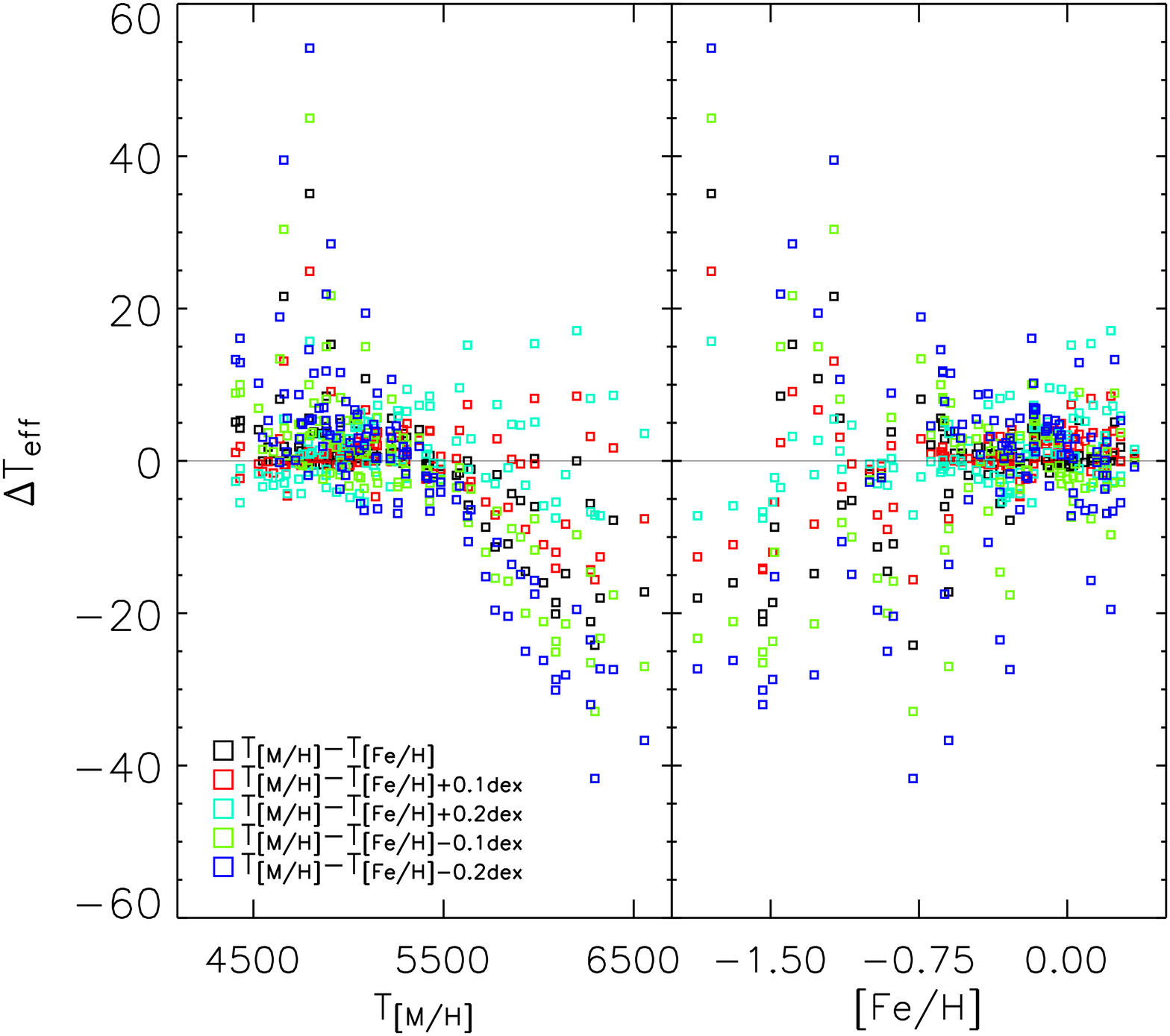}
\caption{{Effect of errors in metallicities on the resulting temperatures.}}
\label{Merr}
\end{center}
\end{figure}

Spectroscopic metallicities have typical errors well within 0.10 dex. As we
show in Figure \ref{Merr} a change of few dex in metallicity has almost no
effect on the resulting temperatures around the solar values, but at lower
metallicities systematic errors of 0.10--0.20 dex in [Fe/H] can introduce
biases up to 40~K. This is an important point when using large surveys with
photometric metallicities that have typical errors above 0.10 dex.

The dependence on the adopted surface gravity is very mild for dwarf stars. We
have verified that a change of $\pm 0.5$ dex in $\log (g)$ implies differences
that never exceed $\sim 30$ K in temperature and well within 1\% for bolometric
fluxes and angular diameters.

\subsubsection{Systematic error in the absolute calibration}
\label{AbsCalSys}

The errors in the adopted magnitudes and absolute calibrations of Vega in
different bands can be regarded as a systematic bias, since once they are
selected, the recovered temperatures, luminosities and angular diameters scale
accordingly. Errors in the observed $BV(RI)_{C}$ magnitudes of Vega are around
0.01 mag whereas the uncertainties in $JHK_S$ magnitudes given by Cohen et
al. (2003) are within a few millimag. As for the random errors on the
photometric zero-points, it is very likely that uncertainties in Vega's
magnitudes compensate each other.

\begin{figure}
\begin{center}
\includegraphics[scale=0.50]{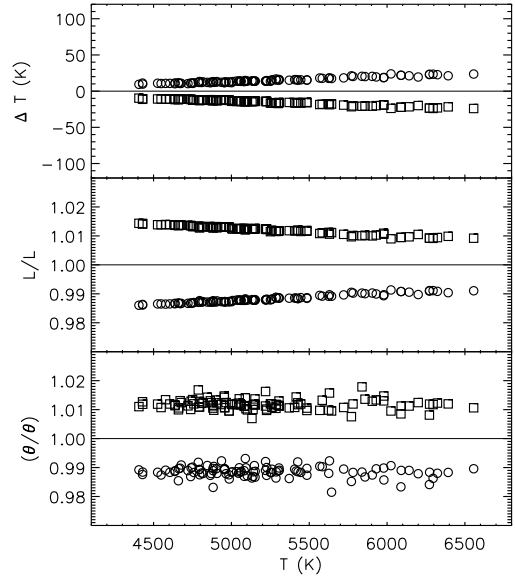}
\caption{{Effect of the absolute calibration being systematically brighter
    (circles) or fainter (squares) when uncertainties given in Table
    \ref{absflux} correlate. Translated into magnitudes it corresponds to a 
    mean shift of Vega's zero-points of $\pm 0.01$ mag in the optical and 
    $\pm 0.02$ mag in the infrared.}}
\label{updown}
\end{center}
\end{figure}

On the other hand, the uncertainties in the adopted absolute calibration mainly
come from the uncertainty in the flux of Vega at 5556 \AA$ $. Since the
absolute calibration in all bands scales accordingly (even though we have used
slightly different approaches in optical and infrared, see Appendix A), we
evaluate the systematic error in temperatures for the worst case scenario,
i.e. when all errors correlate to give systematically higher or lower
fluxes. The uncertainties adopted are those given in Table \ref{absflux}. The
results are shown in Figure \ref{updown}.

Finally, a detailed comparison with data from satellites validate within the
errors the adopted calibration, though it seems to suggest that infrared fluxes
should be brightened by 1\% at most (see Appendix A). If so, the resulting
temperatures would cool down by 10 to 30~K and luminosities and angular
diameters would increase on average by 0.2\% and 0.7\%, respectively.

\subsubsection{The final error budget}

The primary estimate for errors in $T_{\mathrm eff}$ is from the scatter in the
temperature deduced from $J$, $H$ and $K_S$ bands that reflects photometric
errors as well as differences in the corresponding absolute calibration (see
Section \ref{irfmSec}). As regards the uncertainties discussed in Sections
\ref{Felog} and \ref{MC}, they are of the same order and we
quadratically sum them to the standard deviation in the resulting $J$, $H$ and
$K_S$ for a fair estimate of the global errors on temperatures.

The uncertainties of the bolometric luminosities mostly depend on the
photometry that accounts for $\sim$~70--85\% of the resulting luminosity. The
errors from the observations have been estimated by summing in quadrature those
from the MonteCarlo simulation (Section \ref{MC}) to those due to a change of
$\pm 0.5$~dex in $\log (g)$.

Finally, for the resulting angular diameters we propagate the errors in
temperature and luminosity from equation \ref{bolflux}.  The mean internal
accuracy of the resulting temperatures is of 0.8\%, that of the luminosities of
0.5\% and that of the angular diameters of 1.7\%.

There are possible systematic errors coming from the adopted absolute
calibration. As can be seen from Figure \ref{updown} such uncertainties give an
additional mean error of 0.3\% in temperature and 1.2\% in luminosity that
translate into an additional systematic error of 0.7\% in angular diameters. A
summary of the uncertainties is given in Table \ref{unc}.

\begin{table}
\centering
\caption{Mean accuracy of the derived fundamental stellar parameters. Notice
  that either the systematics given here, either those shown in Figure
  \ref{updown} are determined according Table \ref{absflux} and do not account
  for possible (and only indicative) shifts in Vega's zero-points as a
  consequence of its rapidly rotating nature (see Appendix A).}
\label{unc}
\begin{tabular}{ccc}
\hline
                      & Internal accuracy &  Systematics  \\
$T_{\mathrm eff}$     &       0.8\%       &    0.3\%      \\  
$\mathcal{F}_{Bol}$   &       0.5\%       &    1.2\%      \\
$\theta$              &       1.7\%       &    0.7\%      \\
\hline
\end{tabular}
\end{table}

\begin{figure*}
\begin{center}
\includegraphics[scale=0.60]{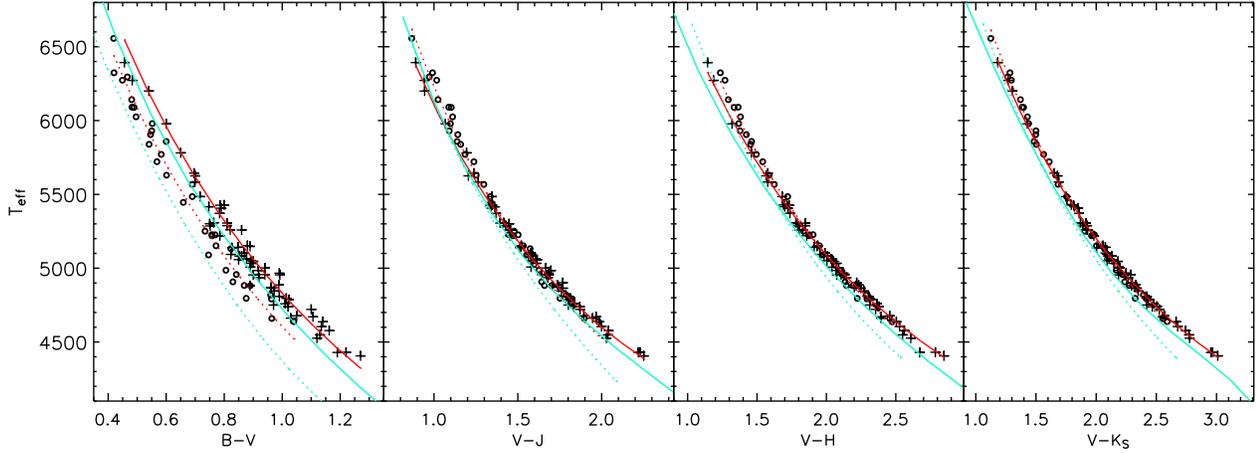}
\caption{{Empirical colour-temperature relations. Circles are for stars with
    [Fe/H] $\le -0.5$, crosses for stars with [Fe/H] $> -0.5$. Red (cyan)
    continuous lines are our (Ram\'irez \& Mel\'endez 2005a) fitting formulae
    at solar metallicity. Dotted lines are for [Fe/H] $=-1$. In the infrared,
    for stars with $T_{\mathrm eff}$ below 5000~K the Ram\'irez \& Mel\'endez
    (2005a) calibration shows a larger metallicity dependence than indicated by
    our data.}}
\label{colteplot}
\end{center}
\end{figure*}

\subsection{Colour-temperature-metallicity fitting formulae}\label{hothot}

To reproduce the observed relation $T_{\mathrm eff}$ vs. colour and to take
into account the effects of different chemical compositions, the following
fitting formula has been adopted (e.g Alonso et al. 1996a, Ram\'irez \& 
Mel\'endez 2005a, Masana et al. 2006):

\begin{displaymath}
\theta_{\mathrm eff}=a_0 + a_1 X + a_2 X^2 + a_3 X \textrm{[Fe/H]} 
\end{displaymath}

\begin{equation}\label{cmt}
\phantom{\theta_{\mathrm eff}=}+ a_4 \textrm{[Fe/H]} + a_5 \textrm{[Fe/H]}^2,
\end{equation} 

where $\theta_{\mathrm eff}=5040/T_{\mathrm eff}$, $X$ represents the colour
and $a_i$ ($i=1,$\ldots,$5$) the coefficients of the fit. In the iterative
fitting, points departing more than $3 \sigma$ from the mean fit were discarded
(very few points ever needed to be removed). All our relations were adequately
fit by simple polynomials, and we did not need to got to higher orders in $X$
to remove possible systematics as function of the metallicity (as in Ram\'irez
\& Mel\'endez, 2005a, who covered a much wider range in temperature than we do
here).

Figure \ref{colteplot} shows the colour-temperature relations in different
bands. The coefficients of the fits, together with the number of stars used,
the range of applicability and the standard deviations are given in Table
\ref{fitcmt}. Note that the very small scatter in the relations reflects the
high-quality and homogeneity of the input data.

\begin{table*}
\centering
\caption{Coefficients and range of applicability of the colour-temperature
  relations.}\label{fitcmt}
\begin{tabular}{lcccccccccc}
Colour & Metallicity range & Colour range & $a_0$ & $a_1$ & $a_2$ & $a_3$ &
$a_4$ & $a_5$ & $N$ & $\sigma (T_{\mathrm eff})$\\
\hline
$B-V$ & $\left[ -1.87, 0.34 \right]$ & $\left[ 0.419, 1.270 \right]$ & 
 0.5121 & 0.5934 & $-$0.0618 & $-$0.0319 & $-$0.0294 & $-$0.0102 & 103 & 54\\
$V-R_C$ & $\left[ -1.87, 0.34 \right]$ & $\left[ 0.249,0.755 \right]$ &
0.4313 & 1.5150 & $-$0.7723 & $-$0.0950 & 0.0179 & $-$0.0033 & 101 & 55\\
$(R-I)_C$ & $\left[ -1.87, 0.34 \right]$ & $\left[0.268, 0.640 \right]$ &
0.2603 &  2.3449 & $-$1.4897 & $-$0.1149 & 0.0641 & 0.0023 & 104 & 70\\
$V-I_C$ & $\left[ -1.87, 0.34 \right]$ & $\left[0.517, 1.395\right]$ &
0.3711 & 0.8994 & $-$0.2467 & $-$0.0545 & 0.0393 & $-$0.0010 & 101 & 49\\
$V-J$ & $\left[ -1.87, 0.34 \right]$ & $\left[0.867, 2.251\right]$ &
0.4613 & 0.4118 & $-$0.0473 & $-$0.0356 & 0.0535 & 0.0012 & 104 & 30\\
$V-H$ & $\left[ -1.87, 0.34 \right]$ & $\left[1.144, 2.847\right]$ &
0.4797 & 0.3059 & $-$0.0252 & $-$0.0196 & 0.0426 & 0.0036 & 99 & 25\\
$V-K_S$ & $\left[ -1.87, 0.34 \right]$ & $\left[1.124, 3.010\right]$ &
0.4609 & 0.3069 & $-$0.0263 & $-$0.0145 & 0.0275 & 0.0006 & 103 & 21\\
\hline
\end{tabular}
\begin{minipage}{1\textwidth}
$N$ is the number of stars employed for the fit after the $3 \sigma$ clipping 
and $\sigma (T_{\mathrm eff})$ the final standard deviation of the proposed
calibrations. 
\end{minipage}
\end{table*}

\section{The temperature scale: some like it hot} 
\label{checking}

In this section we test our IRFM in a number of ways. Firstly, we compare its
predictions to empirical data for the Sun and solar analogs. Secondly, we
compare our predicted angular diameters to recent measurements with large
telescope interferometers for a small sample of G and K dwarfs. Thirdly, we
compare our system to other temperature scales for G and K dwarfs in the
literature.

Overall, we find that our scale is in good agreement with other scales though
some puzzles remain. We make some suggestions for further work.

\subsection{The Sun and solar analogs}
\label{colsu} 

While the temperature, the luminosity and the radius of the Sun are known with
great accuracy, its photometric colours can only be recovered indirectly.
Recently, Holmberg et al. (2006) have provided colour estimates for the Sun
based on those of solar analogs. Another way to obtain colours of the Sun
levers synthetic photometry, convolving empirical or model spectra of the Sun
with filter response functions (e.g. Bessell et al. 1998). Both methods lead to
estimated colours that have typical uncertainties of a few 0.01 mag. Though this
prevents the use of the Sun as a direct calibrator of the temperature scale, it
is still useful to compare to its estimated colours.

\subsubsection{The colours of the Sun}
 
We have computed synthetic solar colours from the solar reference spectrum of
Colina et al. (1996) which combines absolute flux measurements (from satellites
and from the ground) with a model spectrum longward of 9600 \AA. This spectrum,
together with those of three solar analogs, is available from the CALSPEC
library\footnote{ftp://ftp.stsci.edu/cdbs/cdbs2/calspec/}.  According to Colina
et al. (1996), the synthetic optical and near-infrared magnitudes of the
absolutely calibrated solar reference spectrum agree with published values to
within 0.01-0.03 mag.

Recently, two new composite solar spectra extending up to 24000 \AA\ have
been assembled by Thuillier et al. (2004) based on the most recent space-based
data. The accuracy in the UV-visible and near IR is of order 3\%. The two solar
spectra correspond to moderately high and moderately low solar activity
conditions, although the effect of activity on the resulting synthetic colours
is at or below the millimag level, except for $U-B$ (0.003 mag). Such small
differences could be due to the solar spectral variability (which increases
toward shorter wavelengths), the accuracy of the measurements or both. In what
follows we adopt therefore the pragmatic approach of averaging the colours
returned by the two spectra and we generically refer to the results as the
Thuillier et al. (2004) spectrum.

We have also computed the theoretical magnitudes and colours of the Sun
predicted by the latest Kurucz and MARCS synthetic spectra. Magnitudes and
colours computed from the aforementioned solar spectra are given in Table
\ref{sun} together with the empirical colours of Holmberg et al. (2006).

\begin{table*}
\centering
\caption{Magnitudes and colours of the Sun compared to theoretical
colours and those deduced using our temperature scale.}\label{sun}
\begin{tabular}{ccccccccr}
\hline
$V$ & $U-B$ & $B-V$ & $V-R_C$ & $(R-I)_C$ & $V-J$ & $V-H$ & $V-K_S$ & Ref. \\ 
\hline
\ldots & $0.173 \pm 0.064$ & $0.642 \pm 0.016$ &  $0.354 \pm 0.010$  & $0.332 
\pm
0.008$ & $1.151 \pm 0.035$ & $1.409 \pm 0.035$ & $1.505 \pm 0.041$ & (a) \\
$-$26.742 & 0.131 & 0.648 & 0.373 & 0.353 & 1.165 & 1.483 & 1.555 &  (b) \\
$-$26.743 & 0.146 & 0.635 & 0.374 & 0.358 & 1.189 & 1.528 & 1.613 &  (c) \\
$-$26.740 & 0.101 & 0.645 & 0.358 & 0.349 & 1.162 & 1.495 & 1.556 &  (d) \\
$-$26.746 & 0.120 & 0.658 & 0.370 & 0.350 & 1.166 & 1.483 & 1.555 &  (e) \\
$-$26.753 & 0.188 & 0.633 & 0.360 & 0.345 & 1.154 & 1.485 & 1.547 &  (f) \\
\ldots   & \ldots & 0.651 & 0.356 & 0.330 & 1.150 & 1.459 & 1.546 &  (g) \\
\hline
\end{tabular}
\begin{minipage}{1\textwidth}
(a) Holmberg et al. (2006); (b) Colina et al. (1996); (c) Thuillier et
  al. (2004) (d) ATLAS9 ODFNEW with
  Grevesse \& Sauval solar abundances; (e) Kurucz 2004 model with resolving
  power 100000; (f) MARCS; (g) our temperature scale.
\end{minipage}
\end{table*}

Although we have not looked at stellar $U$ magnitudes in this study, for
completeness with the work of Holmberg et al. (2006) we have generated
synthetic magnitudes in this band too. However, in what follows we focus our
discussion from $B$ to $K_S$ band (for a discussion of the theoretical solar
$U-B$ colour see e.g. Grupp 2004b and references therein).  We have used only
Vega to set the zero points. The differences in the resulting colours when also
Sirius is used for the optical bands are given in Appendix A and tend to make
the solar $B-V$ even redder. Assuming no deficiencies in the synthetic spectra,
the uncertainties in the derived colours are entirely due to the uncertainties
in the adopted zero points, i.e. in the order of $\sim$ 0.01 mag.

The CALSPEC and Thuillier et al. (2004) spectra are already absolutely
calibrated for a distance of 1 AU. For the absolute calibration of the Kurucz
and MARCS solar spectra, we also adopt a distance of 1 AU. The synthetic
spectra have a temperature $T_{\odot}=5777\;\textrm{K}$ and we adopt
$L_{\odot}=3.842 \times 10^{33}\,\textrm{erg s}^{-1}$ (Bahcall, Serenelli \&
Basu 2005) from which we deduce $R_{\odot} = \sqrt{L_{\odot}/4 \pi \sigma
T_{\odot}^{4}}$ to be used for the absolute calibration\footnote{The deduced
value $\theta_{\odot}=0.00930179$~rad compares well with that reported in
Landolt-B\"ornstein (1982) $\theta_{\odot}=0.00930484$~rad.}.  Proceeding this
way we can immediately compare the recovered temperatures and luminosities by
means of the IRFM (see Table \ref{temps}) with those given above.  The effect
of the absolute calibration of the solar spectra is immediately seen in the
deduced value of $V$ and reflects in the recovered luminosity whereas the
temperature depends on the colours only.

An influential direct $V$ measurement of the Sun is that of Stebbins \& Kron
(1957) to which Hayes (1985) claimed a further 0.02 mag correction for
horizontal extinction. Using recent standard $V$ magnitudes for the Stebbins \&
Kron (1957) comparison G dwarfs returns for the Sun $V=-26.744\pm0.015$, which
becomes $V=-26.76\pm0.02$ after accounting for the Hayes correction (Bessell,
Castelli \& Plez 1998).  The values deduced using composite and synthetic
spectra are closer to the measurements of Stebbins \& Kron but fully within the
error bars of the Hayes' value. Note that through the year the variation of the
solar distance due to the ellipticity of the Earth's orbit corresponds to a $V$
difference of $\pm 0.035$ mag.

In the optical bands synthetic and composite colours show a remarkably good
agreement with the recent empirical determination, differences being at most
0.01--0.02 mag and therefore within the error bars. Synthetic and composite
spectra also confirm the redder $B-V$ colour with respect to the determination
of Sekiguchi \& Fukugita (2000) ($B-V = 0.626$) and Ram\'irez \& Mel\'endez
(2005a) ($B-V = 0.619$), though the Thuillier et al. (2004) spectrum is
slightly bluer than the Neckel \& Labs (1984) measurements used in the Colina
et al. (1996) composite spectrum.  Using the empirical value of $B-V=0.642$ in
the Ram\'irez \& Mel\'endez (2005a) colour-temperature relation implies a
temperature of the Sun of 5699 K, therefore suggesting the need of a hotter
temperature scale for this colour. Such an offset is of the same order of those
found in Section \ref{emp} for optical and infrared bands.

In the infrared, the only directly measured spectrum is that of Thuillier et
al. (2004). It is much redder than the empirical determination of Holmberg et
al. (2006) and also the model spectra.  As concerns the other models, $J$ band
shows good agreement with the empirical determination. Though a systematic
offset of $\sim$ 0.01 mag appears, it is fully within the error bars.  The $H$
and $K_S$ bands show larger systematic differences in the order of 0.08 and
0.05 mag respectively as Figure \ref{modRAM} already suggests.  Such large
differences are not easily understood in terms of model failures.
Interestingly, an offset of 0.075 mag in the $H$ band zero-point given by the
2MASS has also been claimed by Masana et al. (2006) to constrain a set of solar
analogs to have just the same temperature of the Sun. The reason of such a
difference remains unclear as the adopted temperature scale, the model and
observational uncertainties all might play a role. Among other reasons we
mention that historically the $H$ filter was not defined by Johnson and its
values are not easily homogenized, so that further uncertainties might be
introduced when comparing to various sources, though 2MASS now allows one to
cope with this problem. According to Colina et al. (1996) the IR colours
deduced from the composite spectrum agree within 0.01 mag with the solar
analogs measurements of Wamsteker (1981) and within 0.03 mag with the solar
analogs measurements of Campins et al. (1985). The synthetic colours obtained
using the Colina et al. (1996) solar spectrum also agree very closely with
those obtained when the latest model atmospheres are used.

Using the colours in Table \ref{sun} the temperature of the synthetic Kurucz
and MARCS spectra are always recovered within 7-17~K (see Table \ref{temps}),
differences with the underlying values likely due to the fact that synthetic
solar spectra are tailored to match the observed abundances, $\log(g)$ and
turbulent velocity whereas the interpolation when applying the IRFM is done
over a more generic grid.  Also the temperature of 5802~K deduced when using
the synthetic magnitudes from Colina et al. (1996) solar spectrum agrees very
well with the known value of 5777~K whereas the cooler temperature returned by
the Thuillier et al. (2004) colours is a direct consequence of the much redder
IR colours.  The temperature of the Sun obtained when using the empirical
colours of Holmberg et al. (2006) is 87~K hotter. Changing the empirical $V-H$
from 1.409 to 1.480 lowers the recovered temperature to 5811~K; when also
$V-K_S$ is reddened from 1.505 to 1.550 the recovered temperature then goes to
5774~K. Therefore -- as expected -- the temperature recovered critically
depends on the IR magnitudes in $H$ and $K_S$ bands.  The $H$ band is known to
be troublesome in both spectral modelling and observations, but a discrepancy
in the models $\sim 0.08$ mag is difficult to understand in light of the
comparison with the observations in Section \ref{Compa}. 

\begin{table}
\begin{center}
\caption{Recovered temperature and luminosity of the Sun.}\label{temps}
\begin{tabular}{ccr}
\hline
$T_{\mathrm eff}$~(K) & Luminosity ($\frac{L}{L_{\odot}}$) &  Ref.\\
\hline
5864 & \ldots & (a)\\
5802 & 1.004  & (b)\\ 
5720 & 1.008  & (c)\\
5770 & 0.992  & (d)\\
5794 & 1.003  & (e)\\
5791 & 1.005  & (f)\\
\hline
\end{tabular}
\end{center}
Table Notes: (a) Holmberg et al. (2006); (b) Colina et al. (1996); (c)
  Thuillier et al. (2004); (d) ATLAS9
  ODFNEW with Grevesse \& Sauval solar abundances; (e) Kurucz 2004 model with
  resolving power 100000; (f) MARCS.
\end{table}

The empirical colours of the Sun of Holmberg et al. (2006) are deduced
interpolating in temperature and metallicity a sample of Sun-like stars with
temperatures from Ram\'irez \& Mel\'endez (2005b). This scale is some 100~K
cooler than our own and such a difference can easily account for the
differences in the IR colours. In comparison, the optical colours are almost
unaffected by the temperature used for the fit since they are primarily
dependent on the adopted metallicity to fit the solar analogs.

\subsubsection{The solar analogs}

Absolutely calibrated composite spectra of three solar analogs, namely P041C,
P177D and P330E (Colina \& Bohlin 1997; Bohlin, Dickinson \& Calzetti 2001)
constitute a set of secondary flux standards in addition to the white dwarf
stars with pure hydrogen atmospheres adopted for the HST UV and optical absolute
calibration.  The three solar analogs also determine the absolute flux
distribution of the NICMOS filters.

The solar analogs' flux distribution in the ultraviolet and optical regions is
based on HST \emph{Faint Object Spectrograph} (FOS) and \emph{Space Telescope
Imaging Spectrograph} (STIS) measurements, whereas longward of 10020 \AA$ $ a
scaled version of the Colina et al. (1996) absolute flux of the Sun is used.

The STIS solar analogs' flux distribution in the optical and longer wavelengths
is expected to have uncertainties within 2\% (Bohlin et al. 2001).  Even if the
near-infrared fluxes of the solar analogs have been constructed from models, a
thorough comparison with the high-accuracy $JHK$ infrared photometry of Persson
et al. (1998) gives confidence in the reliability of the adopted spectra. The
differences between the observed magnitudes and the synthetic ones obtained by
using the spectra of the three solar analogs agree within 0.01 mag in $J$
band. The differences in $K$ band are less than 0.025 mag whereas in $H$ band
the models are from 0.03 to 0.07 mag brighter. Such differences can partly
reflect difficulties in modelling this spectral region; however since such a
problem is also present when $H$ band synthetic magnitudes of the white dwarf
calibrators are studied, an alternative or complementary explanation could
simply arise from errors in the adopted zero point (Bohlin et al. 2001;
Dickinson et al. 2002). For the three solar analogs infrared magnitudes are
also available from 2MASS. The observed 2MASS and model magnitudes for each of
the solar analog are given in Table \ref{SA2MASS}. As done by Bohlin et
al. (2001) for the Persson photometry, the $\Delta m$ values compare the
relative observed and model magnitudes to the same difference for P330E (the
primary NICMOS standard):

\begin{displaymath}
\Delta m = [m_{synth}(\textrm{star})-m_{obs}(\textrm{star})]
\end{displaymath} 
\begin{equation}
\phantom{\Delta m =} - [m_{synth}(\textrm{P330E})-m_{obs}(\textrm{P330E})].
\end{equation} 

The uncertainties in the $\Delta m$ values depend only on the repeatability of
the photometry and not on uncertainties in the absolute calibration of the
infrared photometric systems.

\begin{table*}
\centering
\caption{Synthetic and observed 2MASS photometry of HST Solar Analogs.}
\label{SA2MASS}
\begin{tabular}{cccccccccr}
\hline
 $J$  & $J$(synth) & $H$  & $H$(synth) & $K_S$  & $K_S$(synth) & $\Delta m_J$ &
 $\Delta m_H$ & $\Delta m_K$ & Star \\
\hline
10.864 & 10.869 & 10.592 & 10.552 & 10.526 & 10.479 & $-$0.010 & $-$0.058 & $-$0.005 & P041C\\
12.245 & 12.253 & 11.932 & 11.925 & 11.861 & 11.843 & $-$0.007 & $-$0.025 &    0.024 & P177D\\ 
11.781 & 11.796 & 11.453 & 11.471 & 11.432 & 11.390 & \ldots & \ldots & \ldots &  P330E\\
\hline
\end{tabular}
\begin{minipage}{1\textwidth}
P041C = GSC 4413-304; P177D = GSC 3493-432; P330E = GSC 2581-2323. $\Delta m$
values refer to Bohlin's model$-$2MASS difference relative to P330E, as 
explained in the text.
\end{minipage}
\end{table*}

\begin{table*}
\centering
\caption{Magnitudes, colours and temperatures of the Solar
  Analogs. Temperatures have been deduced using the given colours into the
  IRFM.}
\label{analogs}
\begin{tabular}{cccccccccr}
\hline
$V$ & $U-B$ & $B-V$ & $V-R_C$ & $(R-I)_C$ & $V-J$ & $V-H$ & $V-K_S$ &
$T_{\mathrm eff}$~(K) & Star \\ 
\hline
12.005 & 0.135 & 0.611 & 0.346 & 0.346 & 1.137 & 1.454 & 1.526 & 5828 & P041C\\
13.356 & 0.139 & 0.607 & 0.350 & 0.347 & 1.139 & 1.454 & 1.528 & 5831 & P177D\\
12.917 & 0.055 & 0.604 & 0.353 & 0.352 & 1.148 & 1.463 & 1.538 & 5822 & P330E\\
\hline
\end{tabular}
\begin{minipage}{1\textwidth}
Magnitudes and colours of P177D and P330E have been dereddened with the value
of $E(B-V)$ given in Bohlin et al. (2001) and using the standard extinction law
of O'Donnell (1994) and Cardelli, Clayton \& Mathis (1989) in the optical and
infrared, respectively.  The higher temperatures of these solar analogs are in
agreement with the bluer colours and the recovered temperature of P041C agree
very well with the estimated value of 5900~K given in Colina \& Bohlin (1997).
For the other two solar analogs the reddening is used to adjust the continuum
in the infrared to match the UV-optical observations thus avoiding any need to
estimate the temperature (Bohlin, priv. com.)
\end{minipage}
\end{table*}

The IR region of the solar analogs' spectra is a scaled version of that used
for the Sun and the IR synthetic magnitudes of the solar analogs have been
proved to be reliable.  The colours of the solar analogs are bluer than those
obtained from the solar spectra of Colina, Kurucz and MARCS and they are
consistent with slightly higher temperatures for the solar analogs with respect
to the Sun. Nonetheless, $V-H$ and $V-K_S$ are still redder by $\sim0.05$ and
$\sim 0.03$ mag respectively and the disagreement is even worse if we consider
that we are actually comparing solar analogs with hotter temperatures (and
bluer colours) than the Sun.

Finally, we test the temperature scale of Section \ref{hothot} via 11 excellent
solar analogs for which accurate $B,V,J,H,K_S$ colours are available. Seven
are drawn from the top-ten solar analogs of Soubiran \& Triaud (2004) and 4
from the candidate solar twins of King, Boesgaard \& Schuler (2005). The
results are shown in Table \ref{san}.

\begin{table*}
\centering
\caption{Temperature scale applied to Solar Analogs.}
\label{san}
\begin{tabular}{lcccccccc}
\hline
  & [Fe/H] & $V$ & $B-V$ & $J$ & $H$ & $K_S$ & $T_{\mathrm eff}^{\star}$~(K)
  & $T_{\mathrm eff}$~(K)\\ 
\hline
HD 168009& $-0.04$ & 6.30 & 0.64(1) & 5.12(0) & 4.83(6) & 4.75(6) & 5801 & 5784\\
HD 89269 & $-0.23$ & 6.66 & 0.65(3) & 5.39(4) & 5.07(4) & 5.01(2) & 5674 & 5619\\
HD 47309 & $+0.11$ & 7.60 & 0.67(2) & 6.40(3) & 6.16(1) & 6.08(6) & 5791 & 5758\\
HD 42618 & $-0.16$ & 6.85 & 0.64(2) & 5.70(1) & 5.38(5) & 5.30(1) & 5714 & 5773\\
HD 71148 & $-0.02$ & 6.32 & 0.62(4) & 5.15(8) & 4.87(8) & 4.83(0) & 5756 & 5822\\
HD 186427& $+0.06$ & 6.25 & 0.66(1) & 4.99(3) & 4.69(5) & 4.65(1) & 5753 & 5697\\
HD 10145 & $-0.01$ & 7.70 & 0.69(1) & 6.45(0) & 6.11(2) & 6.06(3) & 5673 & 5616\\
HD 129357& $-0.02$ & 7.81 & 0.64(2) & 6.58(3) & 6.25(7) & 6.19(2) & 5749 & 5686\\
HD 138573& $-0.03$ & 7.22 & 0.65(6) & 6.02(7) & 5.74(2) & 5.66(2) & 5710 & 5739\\
HD 142093& $-0.15$ & 7.31 & 0.61(1) & 6.15(8) & 5.91(0) & 5.82(4) & 5841 & 5850\\
HD 143436& $ 0.00$ & 8.05 & 0.64(3) & 6.88(4) & 6.64(9) & 6.54(1) & 5768 & 5813\\
\hline
\end{tabular}
\begin{minipage}{1\textwidth}
Solar analogs drawn from Soubiran \& Triad (2004) and King, Boesgaard \&
Schuler (2005). [Fe/H] and $T_{\mathrm eff}^{\star}$ are those reported in the
two papers. For the stars from Soubiran \& Triaud (2004) we have adopted the
mean metallicity and effective temperature as given in their table
4. $T_{\mathrm eff}$ is the final effective temperature obtained by averaging
the temperatures given by our $(B-V),\, (V-J),\, (V-H),\, (V-K_S)$ calibration.
For HD 168009 and HD 186427 $(V-J)$ has not been used because of the poor
quality flag associated to $J$ magnitudes.
\end{minipage}
\end{table*}
The agreement between our temperatures and those reported in the two papers is
outstanding, the mean difference being only $\Delta T_{\mathrm eff}=-7 \pm
50$~K our temperatures being only slightly cooler. Also, the mean temperature
of the 11 solar analogs set to 5742~K, thus suggesting that our scale is well
calibrated at the solar value.

\subsection{Angular diameters}
\label{AD}

There are several approaches to deriving effective temperatures of
stars. Except when applied to the Sun, very few of them are genuinely direct
methods which measure effective temperature empirically. 

The \emph{direct} methods rely on the measurement of the angular diameter and
bolometric flux of the star.  These fundamental methods are restricted to a
very few dwarf stars, although interferometric (e.g. Kervella et al. 2004) and
transit (Brown et al. 2001) observations have recently increased the sample.
Nonetheless angular diameters obtained from both of the aforementioned methods
have to be corrected for limb-darkening by using some model. Hence the
procedure is still partly model-dependent. Observational uncertainties stem
from systematic effects related to the atmosphere, the instrument and the
calibrators used (e.g. Mozurkewich et al. 2003).  As to giants and supergiants,
where more measurements are available, the comparison of 41 stars observed on
both NPOI and Mark III optical interferometers has shown an agreement within
0.6\%, but with a rms scatter of 4.0\% (Nordgren, Sudol \& Mozurkewich, 2001).

The different limb-darkened values collected by Kervella et al. (2004) for the
same stars give an idea of the uncertainties that might still hinder progress.
For dwarf stars in particular there are still very few measurements, so that a
large and statistically meaningful comparison can not yet be done.  For
example, the limb-darkened measurement of the dwarf stars HD10700 and HD88230
obtained by Pijpers et al. (2003) and Lane et. al (2001), respectively, are
$\sim 5$\% and $\sim 9$\% smaller than those of Kervella et al. (2004) and that
are ultimately used by Ram\'irez \& Mel\'endez (2005b) to check their
scale. Besides, most of the limb-darkening corrections are done using 1D
atmospheric models that rely on the introduction of adjustable parameters like
the mixing length.  3D models for the K dwarf $\alpha$ Cen B provide a radius
smaller by roughly 1$\sigma_{stat}$ (or 0.3\%) compared with what can be
obtained by 1D models (Bigot et al. 2006).  However for hotter stars the
correction due to 3D analysis is expected to be larger as a consequence of more
efficient convection. Interestingly, using parallax to convert the smaller
angular diameter obtained from 3D models into linear radius returns 0.863
$R_{\odot}$, which is in better agreement with the asteroseismic value of 0.857
$R_{\odot}$ by Th\'evenin et al. (2002).  For Procyon the difference between 1D
and 3D models amounts to roughly 1.6\%, implying a correction to $T_{\mathrm
eff}$ of order 50~K (Allende Prieto et al. 2002).

If temperatures are to be deduced from these diameters, further dependence on
bolometric fluxes, often gathered from a variety of non-homogeneous
determinations in literature, enters the game. All these complications
eventually render `direct' methods to be not as direct, or model-independent,
in the end.  In any case, direct angular diameter measurements are still
limited to the nearest (and brightest) dwarfs with metallicity around solar.

In our case, the comparison with \emph{direct} methods is hampered since the
angular diameter measurements available so far are of nearby and bright dwarfs
that have unreliable or saturated 2MASS photometry. Since we do not have any
star in our sample with direct angular diameter measurements, we can only
perform an indirect comparison by using the stars we have in common with
Ram\'irez \& Mel\'endez (2005b).  For the common stars, our angular diameters
are systematically slightly smaller ($\sim 3$\%) than those found by Ram\'irez
\& Mel\'endez (2005b) which agree well with the direct determinations of
Kervella et al. (2004).  Nevertheless, for individual measurements all
determinations agree within the errors.

Finally, we mention another indirect way to determine angular diameters. This
is provided by spectrophotometric techniques, comparing absolutely calibrated
spectra with model ones via equation (\ref{SP}) (Cohen et al. 1996, 1999).  In
the context of the absolutely calibrated spectra assembled by Cohen and
collaborators (see Appendix A) such a procedure has shown good agreement with
direct angular diameter measurements of giants, even though spectrophotometry
leads to angular diameters systematically smaller by a few percent. Considered
the aforementioned scatter for angular diameters of giants, the difference is
not worrisome. However, since the absolute calibration we have used for the
IRFM is ultimately based on the work of Cohen and collaborators (see Appendix
A), such a difference is interesting in light of our results (see also Section
\ref{rm}).

\section{Comparison with other temperature scales}

Most ways to determine effective temperatures are \emph{indirect} methods and
to different extent all require the introduction of models. Other than via the
IRFM, for the spectral types covered by this study, effective temperatures may
be determined via (1) matching observed and synthetic colours (Masana et
al. 2006), (2) the surface brightness technique (e.g. di Benedetto 1998), (3)
fitting observed spectra with synthetic ones (Valenti \& Fischer 2005), (4)
fitting of the Balmer line profile (e.g. Mashonkina et al. 2003), (5) from
spectroscopic conditions of excitation equilibrium of Fe lines (e.g.  Santos et
al. 2004, 2005) and (6) line depth ratios (e.g. Kovtyukh et al. 2003).

In what follows we compare our results with those obtained via these
\emph{indirect} methods. Only single and non-variable stars with accurate
photometry according to the requirements of Section \ref{sample} and
\ref{photo} have been used for the comparison.

\subsection{Ram\'irez \& Mel\'endez sample}
\label{rm}

The widest application of the IRFM to Pop I and II dwarf stars is that of
Alonso et al. (1996b). Recently, Ram\'irez \& Mel\'endez (2005b) have extended
and improved the metallicities in the sample of Alonso et al. (1996b)
recomputing the temperatures.  As expected, the updated temperatures and
bolometric luminosities do not significantly differ from the original ones,
since the absolute infrared flux calibration (i.e. the basic ingredients of the
IRFM) as well as the bolometric flux calibration used are the same as of Alonso
et al.  (1994, 1995). The difference between the old and new temperature scale
is not significant, though in the new scale dwarf stars are some 40~K cooler.

\begin{figure}
\begin{center}
\includegraphics[scale=0.45]{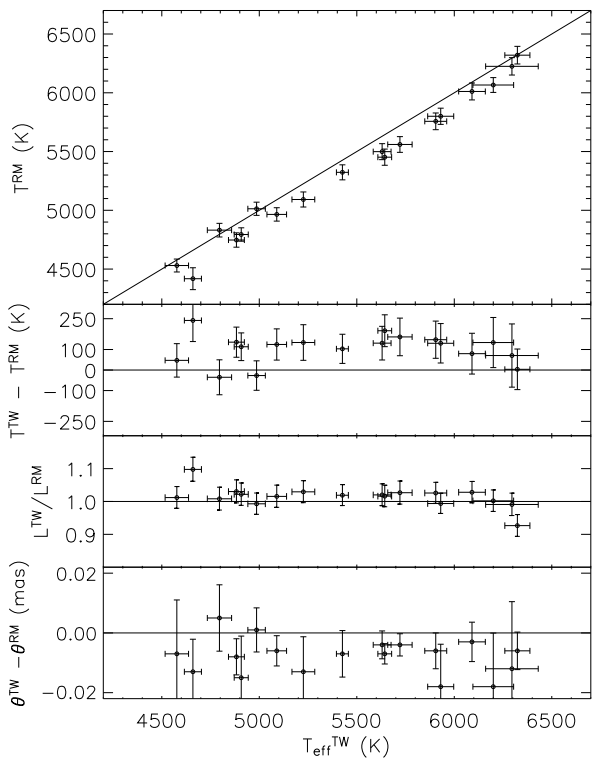}
\caption{{Comparison between effective temperatures, bolometric luminosities
    and angular diameters obtained in this work (TW) and from Ram\'irez \&
    Mel\'endez (RM). We have checked for dependencies in all the scales on
    [Fe/H], $T_{\mathrm eff}, \theta$ and bolometric luminosities and found no
    significant dependencies.}}
\label{ramT}
\end{center}
\end{figure}

For 18 stars in common between our study and that of Ram\'irez \& Mel\'endez
(2005b) we find an average difference $\Delta T_{\mathrm eff}=105 \pm 72$~K,
our scale being hotter. This translates into a mean $T_{\mathrm eff}$
difference of 2.0\%, luminosities brighter by 1.4\% and angular diameters
smaller by 3.3\% (Figure \ref{ramT}). Though not negligible, such differences
are within the error bars of current temperature determinations.

The strength of their temperature scale is that the absolute calibration 
adopted was derived
demanding the IRFM angular diameters to be well scaled to the directly measured
ones for giants (Alonso et al. 1994).  The absolute calibration of Vega 
was tuned to return angular diameters that matched the observed ones; hence it
depends on the input angular diameters and IR photometry that were expected to
be very accurate.  Interestingly, the method applied to hot ($T_{\mathrm eff} >
6000$~K) and cool ($T_{\mathrm eff} < 5000$~K) stars returned absolute 
calibrations
that differed by about 8\% in all IR photometric bands, thus suggesting that
effective temperatures for cool and hot stars and those determined from the
IRFM were not on the same scale.  The cause was some sort of systematic error
affecting direct angular diameters and/or model inaccuracies in predicting IR
fluxes; the final adopted absolute calibration was then a weighted average of
that returned for hot and cool stars.

The comparison of the angular diameters derived by Ram\'irez \& Mel\'endez
(2005b) for dwarf stars with the 13 recent interferometric measurement of
subgiant and main sequence stars (Kervella et al. 2004) seems to imply that the
adopted absolute calibration holds also in this range (Ram\'irez \& Mel\'endez
2004, 2005b).  They find good agreement with direct angular diameter
measurements, but one should be reminded that a well defined angular diameter
scale for dwarf stars is not yet available (see Section \ref{AD}).  Also, the
stars we have in common with Ram\'irez \& Mel\'endez (2005b) have all angular
diameters $\theta < 0.5$ mas and differences between our and their values lie
all below 0.02~mas i.e. below the uncertainties of current measurements (see
table 4 of Ram\'irez \& Mel\'endez 2005b).

We strongly suspect that our hotter temperature scale is the result of the
2MASS vs. Alonso absolute calibration as the almost constant offset in Figure
\ref{ramT} suggests. When we apply our IRFM transforming first 2MASS photometry
into the TCS system used by Alonso, by means of the relations given by
Ram\'irez \& Mel\'endez (2005b) and adopting the TCS filters with the absolute
calibration and Vega's zero-points in the IR given by Alonso, our temperature
scale sets onto that of Ram\'irez \& Mel\'endez within 20~K, with differences
in temperatures, luminosities and angular diameters well below 1\%.  The
confidence in our adopted absolute calibration and zero-points for Vega comes
from the extensive comparison with ground and space based measurements (see
Appendix A). Furthermore, we also prefer to avoid any transformation from the
2MASS to the TCS system since that would introduce further uncertainties of
0.03--0.04 mag in the photometry (Ram\'irez \& Mel\'endez, 2005b).

Finally, we mention another extensive application of the IRFM, that of
Blackwell \& Lynas-Gray (1998). For 10 common stars our mean temperatures are
hotter by $60 \pm 67$~K (1.0\%), our luminosities are brighter by 1.3\% and our
angular diameters smaller by 1.4\%.

\subsection{Masana et al. sample}\label{masana}

Recently Masana et al. (2006) have derived stellar effective temperatures and
bolometric corrections by fitting $V$ and 2MASS IR photometry.  They calibrate
their scale by requiring a set of 50 solar analogs drawn from Cayrel de Strobel
(1996) to have on average the same temperature as the Sun. As a result, they
find significant shifts in the 2MASS zero-points given by Cohen et al. (2003).
Remarkably, the shift in $H$ band is found to be 0.075 mag, i.e. of the same
order of the discrepancy between synthetic and empirical colour (once the
Ram\'irez \& Mel\'endez scale is adopted) found in this band for the Sun (see
Section \ref{colsu}).

\begin{figure}
\begin{center}
\includegraphics[scale=0.45]{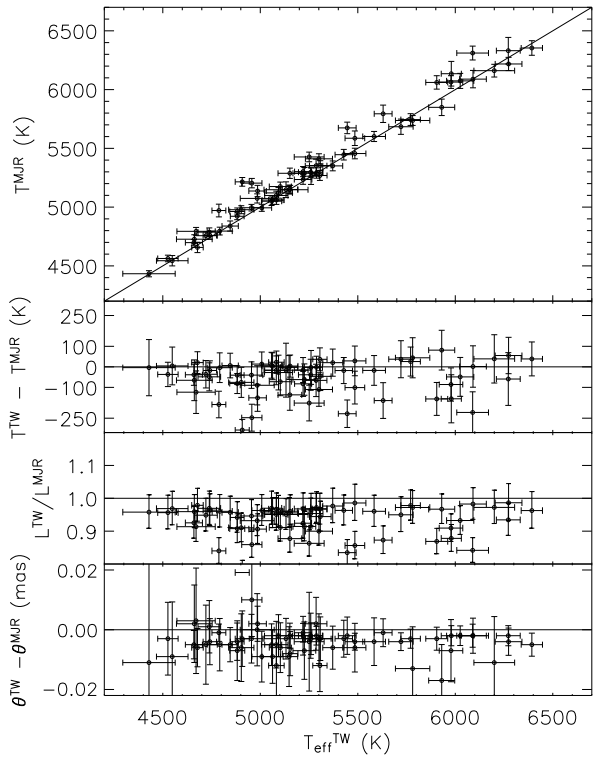}
\caption{{Same as Figure \ref{ramT} but between our work (TW) and the Masana, 
    Jordi \& Ribas scale (MJR).}}
\label{masT}
\end{center}
\end{figure}

For 64 common stars our $T_{\mathrm eff}$ scale is cooler by $50 \pm 80$~K
(1.0\%), our luminosities are fainter by 6.9\% and our angular diameters
smaller by 1.5\% (see Figure \ref{masT}).

Masana et al. (2006) found agreement within 0.3\% between their angular
diameters and the uniform disk measurements collected in the CHARM2 catalogue
of Richichi \& Percheron (2005), with a standard deviation of 4.6\%. Since the
standard deviation is of the same order of the correction between uniform to
limb darkened disk, for the comparison they adopt the questionable choice of
using the uniform disk rather than the proper limb-darkened one.

From 385 stars in common with Ram\'irez \& Mel\'endez (2005b), Masana et al.
(2006) found that their temperature scale is on average 58~K hotter than
Ram\'irez \& Mel\'endez. From stars in common, we find our temperatures are on
average 50~K cooler than that of Masana et al. (2006), leading us to expect we
would be on the same scale as Ram\'irez \& Mel\'endez (2005b), yet from stars
in common (different stars) we find we are 105~K hotter than Ram\'irez \&
Mel\'endez, a very puzzling result! We worked very hard to impute this
inconsistency to one or other of the scales, but were unable to resolve the
problem. This section demonstrates that temperatures, luminosities and angular
diameters calibrations for our stars, retain systematics of the order of a few
percent, despite the high internal accuracy of the data.

\subsection{Di Benedetto sample}

\begin{figure}
\begin{center}
\includegraphics[scale=0.45]{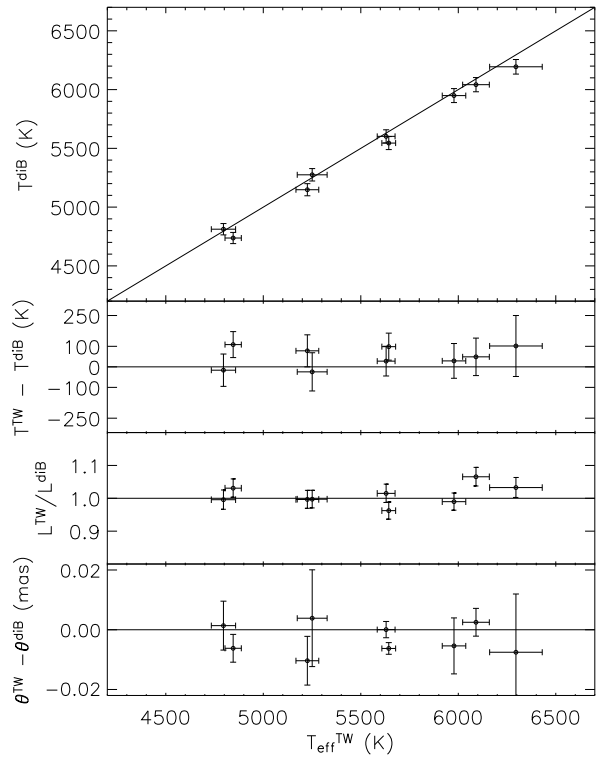}
\caption{{Same as Figure \ref{ramT} but between our work (TW) and the
    Di Benedetto scale (diB).}}
\label{beneT}
\end{center}
\end{figure}

Another way to determine effective temperatures and angular diameters of stars
is via the surface brightness technique. We have 9 stars in common with the
extensive work of Di Benedetto (1998). On average our stars are $50 \pm 50$~K
hotter, the luminosities are brighter by 0.9\% and the angular diameters
smaller by 1.4\% (see Figure \ref{beneT}).

Comparing to the angular diameters from the IRFM sample of Blackwell \& Lynas
Gray (1998), Di Benedetto (1998) found in the F-G-K spectral range an overall
agreement well within 1\% though with an intrinsic scatter as large as 2\%.

\subsection{Valenti \& Fischer sample}

Valenti \& Fischer (2005) have presented a uniform catalogue of stellar
properties for 1040 nearby F, G and K stars that have been observed by the
Keck, Lick and AAT planet search programs. Fitting observed echelle spectra
with synthetic ones, they have obtained effective temperatures, surface
gravities and abundances for every star. We have 47 stars in common. Except for
two of them (HD22879 and HD193901) which depart from the comparison by 242~K
and 497~K respectively, we obtain an excellent average agreement $\Delta
T_{\mathrm eff}=6 \pm 60$ K.  We could not single out a reason of such a large
discrepancy with the two outliers; however we are confident in our values since
they agree within 80-150 K with the temperatures of Ram\'irez \& Mel\'endez
(2005b) and Masana et al. (2006).

\subsection{Santos et al. sample}

Santos et al. (2004, 2005) have carried out a detailed spectroscopic analysis
for 119 planetary-host stars and 95 single stars.  They obtained spectroscopic
temperatures based on the analysis of several Fe I and Fe II lines. The
comparison done by Ram\'irez \& Mel\'endez (2004, 2005b) seems to imply that
such temperatures are hotter by about $\sim 100$~K. However for stars with
direct angular diameter measurements Santos (2005) found an agreement within
7~K with the direct measured temperatures reported in Ram\'irez \& Mel\'endez
(2004).

After removing a star that departs 280~K (HD142709), the difference between our
and Santos $T_{\mathrm eff}$ scale is to $15 \pm 81$~K, our scale being cooler.
The reason of the large departure for the one outlier is not clear, but our
value agree better with Masana et al. (2006).

Another study with temperatures derived using strictly spectroscopic criteria
(based on the excitation equilibrium of the iron lines) is that of Luck \&
Heiter (2005). We have only 5 stars in common, however the average agreement is
good ($\Delta T_{\mathrm eff} = 11 \pm 142$~K). Even though such a result is
dominated by the scatter, it further suggests that our IRFM temperatures agree
very well with spectroscopic determinations.

\subsection{Mashonkina et al. sample}

Balmer line profile fitting allows a very precise determination of stellar
effective temperature for cool stars (Fuhrmann, Axer \& Gehren 1993, 1994).
Mashonkina et al. (2000, 2001, 2003) have extensively used such a technique to
derive effective temperatures to be used for their detailed abundance analysis.
We have 9 stars in common and the mean difference is $37 \pm 64$~K, our
temperatures being hotter.

\subsection{Kovtyukh et al. sample}

The line depth ratios technique (Gray 1989, 1994; Gray \& Johanson 1991)
applied to high $S/N$ echelle spectra is capable of achieving an internal
precision as high as 5 to 10~K. Kovtyukh et al. (2003) has applied such a
technique to a set of 181 F-K dwarfs and adjusted the zero-point of the scale
on the solar reflected spectra taken with ELODIE, leading to the uncertainty in
the zero-point of order 1~K. Unfortunately we only have three stars in common
(in the range 5000 to 5500~K), too few to draw any meaningful conclusions.
Nevertheless we find good agreement with a mean difference of $37 \pm 43$~K,
our stars being cooler. The scale of Kovtyukh is in fact hotter by about 90~K
when compared with the stars in the same range from Ram\'irez \& Mel\'endez
(2005b).  Since the zero-point of Kovtyukh et al. (2003) is calibrated on the
Sun, one of their conclusion is that either the IRFM of Alonso et al. (1996b)
and Blackwell \& Lynas-Gray (1998) or the surface brightness technique of di
Benedetto (1998) could actually predict a too low temperature for the Sun and
the solar type stars. As we have seen in Section \ref{colsu} a hotter
temperature scale for the solar analogs could indeed solve the discrepancies
between observed and computed colours in the infrared.

\subsection{Summary : systematic error remains the problem}

Our temperature scale is slightly hotter than other scales based on the IRFM,
typically by 50 to 100 K. Our scale agrees closely to the temperature scale
established via spectroscopic methods. The strength of our work is that our
IRFM was done completely from first principles (from the multiband photometry
and an adopted absolute flux calibration) with the best quality data
available. The bolometric fluxes of the stars
are close to completely observationally constrained (only 15 to 30 \% of the
stellar flux lies outside our $BV(RI)_C JHK_S$ filters). Despite this effort,
comparison with many other temperature scales forces us to conclude that
external uncertainties in the temperature scale of lower main sequence stars
remains of of order $\pm100$~K. The external uncertainty dominates our high
internal error, which is of order only 40~K.  

A summary of the comparisons of our scale with the others mentioned above is
given in Table \ref{temp}.

\begin{table*}
\centering
\caption{Comparison with other works.}\label{temp}
\begin{tabular}{lccccccccc}
 & Ram\'irez \& & Blackwell \& & Masana & Di Benedetto & Valenti \& &  Santos & 
Luck \& &  Mashonkina & Kovtyukh\\
 & Mel\'endez   & Lynas-Gray   & et al. &              & Fischer    &   et al. & 
Heiter  &      et al. & at al.\\
\hline
Method              & (a) & (a) & (b)   & (c)& (d)   & (e)   & (e)   & (f)  & (g) \\
Stars in common     & 18  & 10  &  64   & 9  & 45 (2)& 14 (1)& 5     & 9    & 3  \\
$\Delta T_{\mathrm eff}$~(K)& 105 & 60  & $-$50 &50  & 6 & $-$15 & $-$11 & 37   & $-$37\\
$\sigma$~(K)        & 72  & 67  &  80   &50  & 60    &  81   & 142   & 64   & 43 \\
$\Delta L$~(\%)     & 1.4 & 1.3 &$-$6.9\%& 0.9\% & \ldots & \ldots & \ldots &
   \ldots & \ldots \\ 
$\Delta \theta$~(\%)& $-$3.3 & $-$1.4 & $-$1.5 & $-$1.4 & \ldots & \ldots & \ldots &
   \ldots & \ldots \\ 
\hline
\end{tabular}
\begin{minipage}{1\textwidth}
(a) = IRFM; (b) = Spectral Energy Distribution fit; (c) = Surface Brightness
  technique; (d) =  fitting observed spectra with synthetic ones; (e) =
  excitation equilibrium of Fe lines; (f) = fitting of the Balmer line
  profile; (g) = line depth ratios. The number of stars in the bracket refers
  to those not counted because of large departures in temperature. Differences
  are computed our work$-$others.
\end{minipage}
\end{table*}

\section{Empirical bolometric corrections}\label{BolSec}

In this section we derive bolometric corrections for our stars. The definition
of apparent bolometric magnitude is:

\begin{equation}
m_{Bol} = -2.5 \log ( \mathcal{F}_{Bol} ) + \textrm{constant}
\end{equation}

where $\mathcal{F}_{Bol}$ is the bolometric flux received on the Earth as
defined in eq. (\ref{bolflux}).  The usual definition of bolometric correction
in a given band

\begin{equation}\label{bolcor}
\textrm{BC}_{\zeta} = m_{Bol}-m_{\zeta}=M_{Bol}-M_{\zeta}
\end{equation}

where BC is to be added to the magnitude in a given band $\zeta$ to yield the
bolometric magnitude. 

Although the definition of bolometric magnitude is straightforward, there can
be some confusion resulting from the choice of zero-point. Originally
bolometric corrections where defined for the $V$ band only and it had been
accepted that $BC_{V}$ should be negative for all stars. From the spectral
energy distribution we expect the bolometric correction in the $V$ band to be
largest for very hot and for very cool stars. The minimum of $BC_V$ then occurs
around spectral type F, which then set the zero-point of the bolometric
magnitudes. This implied a value $BC_V$ for the Sun of between $-$0.11 mag
(Aller, 1963) and $-$0.07 (Morton \& Adams, 1968). However with the publication
of a larger grid of model atmospheres, the smallest bolometric correction in
Kurucz's grid (1979) implied $BC_V=-0.194$. The original choice of the
zero-point for the bolometric corrections has thus proved troublesome. A better
method is to adopt a fixed zero point.

We \emph{define} the absolute bolometric magnitude of the Sun to be
$M_{Bol,\odot}=4.74$ in accordance with Bessell et al. (1998). By adopting its
measured apparent magnitude $V_{\odot}=-26.76$ (see Section \ref{colsu}), the
absolute $M_V$ magnitude is thus 4.81 and the $V$ bolometric correction is then
$BC_V=4.74-4.81=-0.07$.

The absolute bolometric magnitude for a star with luminosity $L$, radius $R$
and effective temperature $T_{\mathrm eff}$ is

\begin{displaymath}
M_{Bol}=-2.5 \log \left(L/L_{\odot} \right) + M_{Bol,\odot}
\end{displaymath}
\begin{equation}\label{bolma}
\phantom{M_{Bol}}=-2.5 \log \left( R^2 T_{\mathrm eff}^4 / R_{\odot}^2
T_{\mathrm eff,\odot}^4 \right) + M_{Bol,\odot}
\end{equation}

where $L=4 \pi R^2 \sigma T_{\mathrm eff}^4$ and $L_{\odot}=3.842 \times
10^{33}$ erg $\textrm{s}^{-1}$ (Bahcall et al. 2005).

Using eq. (\ref{bolma}) and the definition of absolute magnitude
$M_{\zeta}=m_{\zeta}+5 \log (\pi) - 10$, where $\pi$ is the parallax (in mas),
the bolometric correction (\ref{bolcor}) is

\begin{displaymath}
BC_{\zeta}= -5 \log \frac{R \pi}{R_{\odot}}-10 \log
\frac{T_{\mathrm eff}}{T_{\mathrm eff,\odot}}+M_{Bol,\odot}-m_{\zeta}+10
\end{displaymath}

\begin{equation}\label{tricky}
\phantom{BC_{\zeta}}= -5 \log \mathcal{K} \frac{\theta/2}{R_{\odot}}-10 \log
\frac{T_{\mathrm eff}}{T_{\mathrm eff,\odot}}+M_{Bol,\odot}-m_{\zeta}+10
\end{equation}

where $\mathcal{K}$ is the conversion factor for the proper unit
transformation, $\theta = 2R\pi / \mathcal{K}$ is the angular diameter (in mas)
determined via IRFM and $R_{\odot}$ is the solar radius in cm as given in
Section \ref{colsu}. From grids of model atmospheres bolometric corrections can
be calculated in a similar manner, where the dependence on the radius of the
star eliminates in the difference between the absolute bolometric and in-band
magnitudes (e.g. Girardi et al. 2002).

\begin{figure*}
\begin{center}
\includegraphics[scale=0.87]{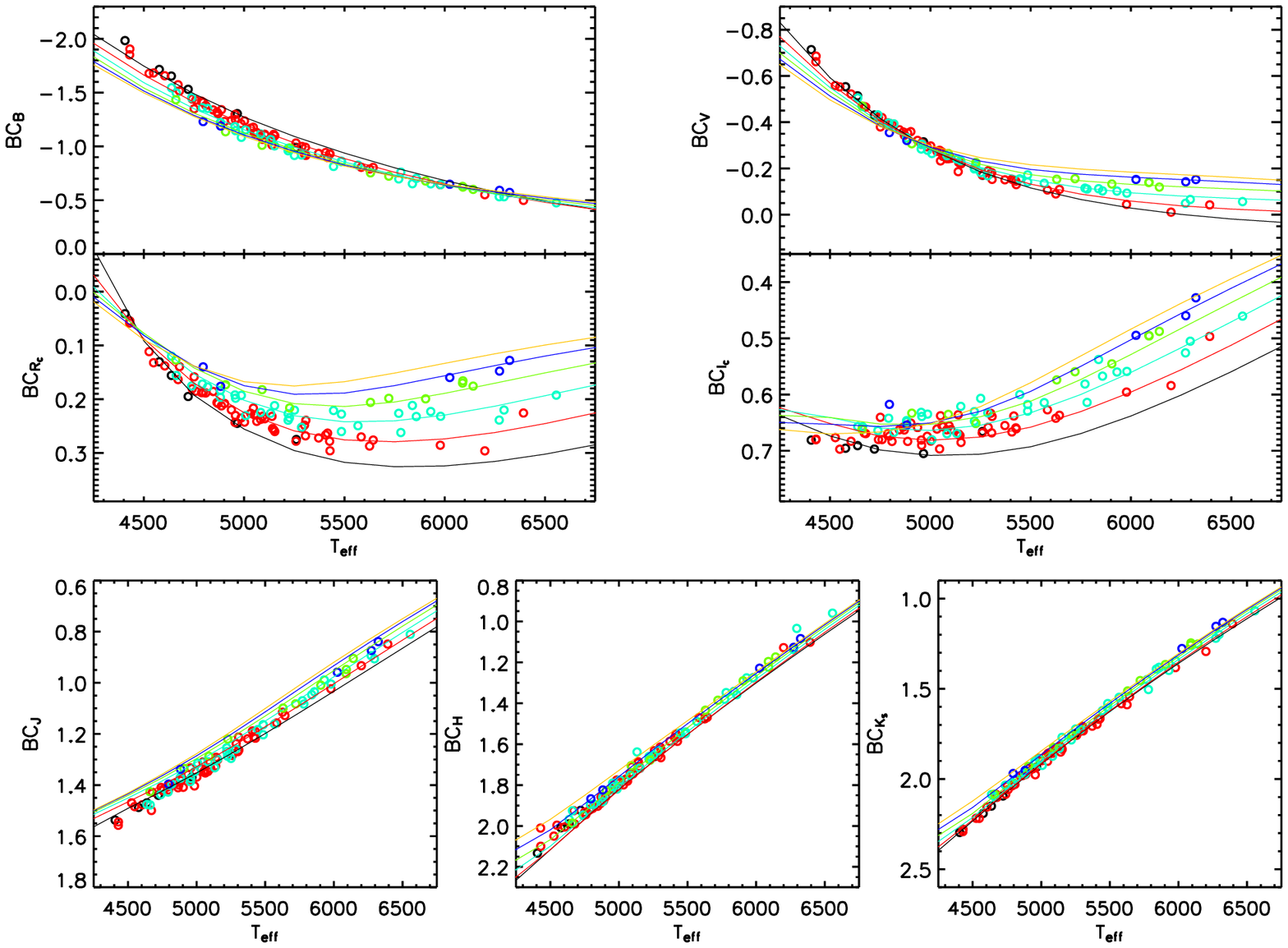}
\caption{{Bolometric corrections calculated from Castelli \& Kurucz (2003)
ATLAS9 are represented as function of $T_{\mathrm eff}$. [M/H] equal to +0.5 (black
line), +0.0 (red line), $-$0.5 (cyan line), $-$1.0 (green line), $-$1.5 (blue
line), $-$2.0 (yellow line). Points correspond to our IRFM bolometric
corrections for the
sample stars in the range [M/H] $> 0.25$ (black), $-0.25 <$ [M/H] $\leq$ 0.25
(red), $-0.75 <$ [M/H] $\leq -0.25$ (cyan), $-1.25 <$ [M/H] $\leq -0.75$
(green), $-1.75 <$ [M/H] $\leq -1.25$ (blue). The metallicities given for the
model are solar-scaled, whereas [M/H] for the stars has been computed using
eq. (\ref{overme}).}}
\label{theo}
\end{center}
\end{figure*}

The comparison between the empirical bolometric corrections in different bands
computed via equation (\ref{tricky}) and those predicted by model atmospheres
(ATLAS9) is shown in Figure \ref{theo}.

This plot can be regarded as the theoretical counterpart of Figure
\ref{KurCol}.  Again, the agreement between model atmospheres and empirical
data is very good and it should be remembered that the model dependence in
deducing empirical bolometric correction from our implementation of the IRFM 
is small (only few 10 \%, see Section \ref{irfm}).

A complementary way of deriving stellar integrated fluxes is via photometric
indices. The integrated flux of a star, $\mathcal{F}_{Bol}\textrm{(Earth)}$,
depends primarily on its apparent brightness (especially in $R_C$ and $I_C$
bands, see later), which may be measured by its magnitude in different
bands. Of lesser importance is its temperature, which is a function of a colour
index and metallicity, $\phi (X,\textrm{[Fe/H]})$.  Following Blackwell \&
Petford (1991a), we expect a relation of the form

\begin{equation}
\mathcal{F}_{Bol}\textrm{(Earth)}=10^{-0.4 m_{\zeta}} \phi(X,\textrm{[Fe/H]}).
\end{equation}

\begin{figure*}
\begin{center}
\includegraphics[scale=0.74]{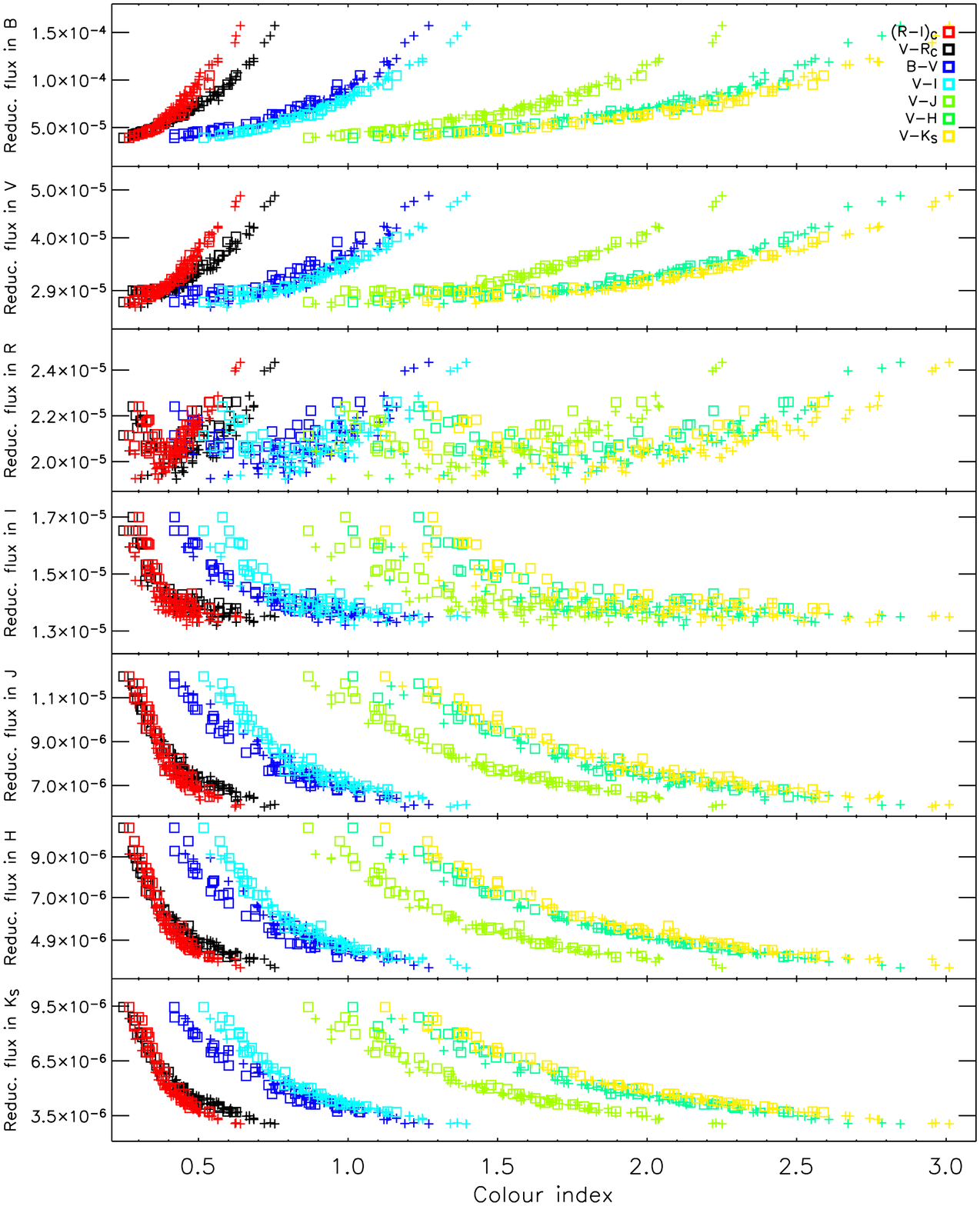}
\caption{{Reduced flux in different $\zeta$-bands ($\mathcal
{F}_{Bol}\textrm{(Earth)}\,10^{0.4\, m_{\zeta}}$) plotted as function of
different colour indices. Squares are for stars with [Fe/H]~$\le -0.5$, crosses
for stars with [Fe/H]~$> -0.5$.}}
\label{phi}
\end{center}
\end{figure*}

The function $\phi (X,\textrm{[Fe/H]})$ is illustrated in Figure \ref{phi}, in
which the reduced flux in $\zeta$-band, $\mathcal
{F}_{Bol}\textrm{(Earth)}\,10^{0.4\, m_{\zeta}}$, is plotted against different
colour indices. For the integrated stellar flux we have fitted expressions of
the form:

\begin{displaymath}
\mathcal{F}_{Bol}\textrm{(Earth)}=10^{-0.4 m_{\zeta}} \Big[ b_0+b_1 X + 
         b_2 X^2 + b_3 X^3 
\end{displaymath}
\begin{equation}
\phantom{\mathcal{F}_{Bol}\textrm{(Earth)}=\Big[ } + 
 b_4 X  \textrm{[Fe/H]} + b_5 \textrm{[Fe/H]} + b_6 \textrm{[Fe/H]}^2 \Big].
\end{equation}

We tried different fitting formulae but this form proved the most
satisfactory. There are a few fits of this type in the literature (Blackwell \&
Petford 1991a for a subset of the colours used here, and Alonso et al.  1995
for the infrared). Blackwell et al's fit has no metallicity dependence because
metallicities were not available. Alonso et al's formula gives fits which are
good in the IR but do not reproduce the observed trend in the optical bands.
Our fitting formulae work well from optical to IR.  As for the
temperature-colour fitting, at every iteration points departing more than $3
\sigma$ from the mean fit were discarded. The coefficients of the fits,
together with the number of stars used, the range of applicability and the
standard deviation of the differences between the measured fluxes and the
fluxes calculated from the fitting formula are given in Table \ref{bolfit}.

\begin{table*}
\centering
\caption{Coefficients and range of applicability of the absolute flux 
         calibrations. All calibrations are valid for the metallicity range
         $-1.87 \le \textrm{[Fe/H]} \le 0.34$.}\label{bolfit}
\begin{tabular}{cccccccccccc}
Reduced & Colour & Colour & $b_0$ & $b_1$ & $b_2$ 
& $b_3$ & $b_4$ & $b_5$ & $b_6$ & $N$ & $\sigma(\%)$\\
flux & & range &  & & & & & & & & \\
\hline
$B$ & $B-V$   & $\left[0.419, 1.270\right]$ & 
1.6521 & 8.6205 & $-$13.7586 & 12.5042 & $-$0.7734 & $-$0.0644 & $-$0.0882 & 100 &1.7\\
    & $V-R_C$   & $\left[0.249, 0.755\right]$ & 
0.4475 & 19.0419 & $-$32.6586 & 44.0700 & $-$0.8436 & 0.4652 & 0.1417 & 98 & 2.3\\
  & $(R-I)_C$   &  $\left[0.268, 0.640\right]$ & 
11.3466 & $-$69.2615 & 184.7907 & $-$105.3337 & 4.8308 & $-$1.1543 & 0.1516 & 101 &
3.5\\ 
   & $V-I_C$   & $\left[0.517, 1.395\right]$ & 
3.9334 & $-$4.2082 & 7.7254 & 0.7865 & 0.5466 & $-$0.1401 & 0.1045 & 99 & 1.8\\
   & $V-J$   & $\left[0.867, 2.251\right]$ & 
-0.5405 & 9.0544 & $-$6.6975 & 2.5713 & 0.5042 & $-$0.2272 & 0.1393 & 99 & 2.5\\
   & $V-H$   &  $\left[1.016, 2.847\right]$ & 
-1.3883 & 8.9784 & $-$5.3419 & 1.4980 & 0.5184 & $-$0.3228 & 0.1759 & 99 & 2.4\\
   & $V-K_S$   &  $\left[1.124, 3.010\right]$ & 
-0.3850 & 6.4601 & $-$3.5378 & 1.0324 & 0.4693 & $-$0.4049 & 0.1425 & 102 & 2.0\\
 & & & & & & & & & & & \\
$V$ & $B-V$   &  $\left[0.419, 1.270\right]$ & 
2.9901 & $-$1.9408 & 1.6146 & 0.9227 & $-$0.0857 & $-$0.3010 & $-$0.0800 & 96 & 1.5\\
   & $V-R_C$   &  $\left[0.249, 0.755\right]$ &
2.4783 & $-$0.6395 & 2.4817 & 3.3947 & $-$0.0222 & $-$0.1769 & $-$0.0100 & 103 & 1.2\\
  & $(R-I)_C$   &  $\left[0.268, 0.640\right]$ & 
6.0078 & $-$28.2448 & 69.7676 & $-$44.6415 & 1.2198 & $-$0.5151 & 0.0052 & 99 &
1.7\\
  & $V-I_C$   &  $\left[0.517, 1.395\right]$ & 
3.7933 & $-$5.2461 & 6.3043 & $-$1.4363 & 0.1987 & $-$0.2959 & $-$0.0118 & 100 & 0.9\\
   & $V-J$   &  $\left[0.867, 2.251\right]$ & 
2.9726 & $-$0.9034 & 0.3570 & 0.1879 & 0.1641 & $-$0.3144 & 0.0062 & 96 & 0.9\\
  & $V-H$   &  $\left[1.016, 2.847\right]$ & 
2.5911 & $-$0.0458 & $-$0.1431 & 0.1563 & 0.1532 & $-$0.3338 & 0.0121 & 99 & 0.9\\
   & $V-K_S$   &  $\left[1.124, 3.010\right]$ & 
3.0690 & $-$0.8886 & 0.3413 & 0.0497 & 0.1410 & $-$0.3528 & 0.0024 & 102 & 0.9\\
 & & & & & & & & & & & \\
$R$ & $B-V$   &  $\left[0.419, 1.270\right]$ & 
2.6213 & $-$2.0936 & 1.6873 & $-$0.1037 & $-$0.0267 & $-$0.1361 & $-$0.0229 & 100 &
1.0\\
   & $V-R_C$   &  $\left[0.249, 0.755\right]$ &
2.4703 & $-$2.9446 & 4.5916 & $-$1.0001 & 0.0889 & $-$0.1652 & $-$0.0075 & 103 & 1.2\\
  & $(R-I)_C$   & $\left[0.268, 0.640\right]$ & 
3.7090 & $-$12.3326 & 26.6386 & $-$16.4346 & 0.5423 & $-$0.3234 & $-$0.0055 & 101 &
0.9\\ 
  & $V-I_C$   &  $\left[0.517, 1.395\right]$ & 
2.9332 & $-$3.1362 & 2.9790 & $-$0.7098 & 0.1182 & $-$0.2203 & $-$0.0090 & 103 & 1.0\\
   & $V-J$   &  $\left[0.867, 2.251\right]$ & 
2.9077 & $-$1.7416 & 0.9045 & $-$0.1007 & 0.0983 & $-$0.2532 & $-$0.0046 & 95 & 0.6\\
  & $V-H$   &  $\left[1.016, 2.847\right]$ & 
2.6787 & $-$0.9882 & 0.3538 & $-$0.0127 & 0.0790 & $-$0.2506 & $-$0.0041 & 91 & 0.6\\
   & $V-K_S$   &  $\left[1.124, 3.010\right]$ & 
2.8723 & $-$1.2499 & 0.4812 & $-$0.0383 & 0.0689 & $-$0.2430 & $-$0.0070 & 93 & 0.6\\
 & & & & & & & & & & & \\
$I$ & $B-V$   &  $\left[0.419, 1.270\right]$ & 
2.6225 & $-$3.6174 & 3.4760 & $-$1.1216 & 0.0101 & $-$0.0438 & $-$0.0060 & 98 & 1.0\\
  & $V-R_C$   &  $\left[0.249, 0.755\right]$ &
2.7401 & $-$7.1587 & 12.3123 & $-$7.0236 & 0.1591 & $-$0.1251 & $-$0.0077 & 100 &
1.0\\
  & $(R-I)_C$   & $\left[0.268, 0.640\right]$ &
3.2531 & $-$11.0246 & 21.2495 & $-$13.5880 & 0.4144 & $-$0.2412 & $-$0.0038 & 102 &
1.0\\ 
  & $V-I_C$   &  $\left[0.517, 1.395\right]$ & 
2.9439 & $-$4.3412 & 3.9408 & $-$1.1868 & 0.1287 & $-$0.1717 & $-$0.0061 & 103 & 1.0\\
   & $V-J$   &  $\left[0.867, 2.251\right]$ & 
2.9412 & $-$2.6212 & 1.4468 & $-$0.2658 & 0.0881 & $-$0.1967 & $-$0.0077 & 97 & 1.1\\
  & $V-H$   &  $\left[1.016, 2.847\right]$ & 
2.8196 & $-$1.8941 & 0.8205 & $-$0.1187 & 0.0645 & $-$0.1866 & $-$0.0077 & 96 & 1.1\\
   & $V-K_S$   &  $\left[1.124, 3.010\right]$ & 
2.8475 & $-$1.8308 & 0.7520 & $-$0.1031 & 0.0540 & $-$0.1688 & $-$0.0078 & 100 & 1.0\\
 & & & & & & & & & & & \\
$J$ & $B-V$   &  $\left[0.419, 1.270\right]$ &
2.4518 & $-$4.0671 & 3.2357 & $-$0.9315 & $-$0.0658 & 0.1077 & 0.0185 & 93 & 2.2\\
  & $V-R_C$   &  $\left[0.249, 0.755\right]$ &
2.8867 & $-$10.0254 & 15.7145 & $-$8.5556 & 0.0763 & $-$0.0217 & 0.0067 & 93 & 2.0\\
  & $(R-I)_C$   &  $\left[0.268, 0.640\right]$ &
3.7432 & $-$15.8775 & 27.7771 & $-$16.6189 & 0.3220 & $-$0.1709 & 0.0065 & 94 &
2.5\\
  & $V-I_C$   &  $\left[0.517, 1.395\right]$ & 
3.2376 & $-$6.2189 & 5.1663 & $-$1.4808 & 0.0931 & $-$0.0901 & 0.0061 & 90 & 2.1\\
   & $V-J$   &  $\left[0.867, 2.251\right]$ & 
2.7711 & $-$2.8212 & 1.3163 & $-$0.2179 & 0.0628 & $-$0.1140 & 0.0022 & 96 & 0.9\\
  & $V-H$   &  $\left[1.144, 2.847\right]$ & 
2.8830 & $-$2.4146 & 0.9164 & $-$0.1227 & 0.0382 & $-$0.0962 & 0.0003 & 91 & 1.7\\
   & $V-K_S$   & $\left[1.124, 3.010\right]$ & 
2.7876 & $-$2.1525 & 0.7601 & $-$0.0949 & 0.0294 & $-$0.0732 & 0.0039 & 90 & 1.5\\
 & & & & & & & & & & & \\
$H$ & $B-V$   & $\left[0.419, 1.270\right]$ &
2.3593 & $-$4.3075 & 3.3103 & $-$0.9179 & $-$0.1030 & 0.1659 & 0.0263 & 97 & 2.9\\
  & $V-R_C$   &  $\left[0.249, 0.755\right]$ &
2.7616 & $-$10.2963 & 15.6654 & $-$8.2985 & 0.0264 & 0.0186 & 0.0097 & 95 & 2.8\\
  & $(R-I)_C$   &  $\left[0.268, 0.640\right]$ &
3.5683 & $-$15.8279 & 26.9524 & $-$15.7470 & 0.3107 & $-$0.1504 & 0.0105 & 95 & 3.2\\
  & $V-I_C$   &  $\left[0.517, 1.395\right]$ &
3.0862 & $-$6.2854 & 5.0826 & $-$1.4230 & 0.0780 & $-$0.0624 & 0.0083 & 95 & 2.8\\
   & $V-J$   &  $\left[0.867, 2.251\right]$ & 
2.6182 & $-$2.8623 & 1.2906 & $-$0.2070 & 0.0518 & $-$0.0823 & 0.0053 & 99 & 2.5\\
  & $V-H$   &  $\left[1.016, 2.847\right]$ & 
2.4555 & $-$2.0593 & 0.7245 & $-$0.0916 & 0.0386 & $-$0.0839 & 0.0028 & 100 & 1.0\\
   & $V-K_S$   &  $\left[1.124, 3.010\right]$ &
2.5220 & $-$2.0243 & 0.6752 & $-$0.0804 & 0.0246 & $-$0.0505 & 0.0049 & 97 & 2.0\\
 & & & & & & & & & & & \\
$K_S$ & $B-V$   &  $\left[0.419, 1.270\right]$ &
2.2158 & $-$4.0992 & 3.1957 & $-$0.9165 & $-$0.0680 & 0.1228 & 0.0208 & 102 & 2.8\\
  & $V-R_C$   &  $\left[0.249, 0.755\right]$ &
2.6509 & $-$10.0609 & 15.4526 & $-$8.3102 & 0.0587 & $-$0.0096 & 0.0049 & 97 & 2.9\\
  & $(R-I)_C$   &  $\left[0.268, 0.640\right]$ &
 3.2130 & $-$13.9314 & 23.1287 & $-$13.2435 & 0.3450 & $-$0.1801 & 0.0049 & 100 &
 2.9\\ 
  & $V-I_C$   &  $\left[0.517, 1.395\right]$ &
2.8573 & $-$5.7901 & 4.6467 & $-$1.3010 & 0.0958 & $-$0.0924 & 0.0028 & 103 & 2.7\\
   & $V-J$   &  $\left[0.867, 2.251\right]$ & 
2.5152 & $-$2.8220 & 1.2970 &  $-$0.2136 & 0.0645 & $-$0.1172 & $-$0.0001 & 100 &
2.6\\
  & $V-H$   &  $\left[1.016, 2.847\right]$ & 
2.4184 & $-$2.1148 & 0.7671 & $-$0.1005 & 0.0453 & $-$0.1096 & $-$0.0014 & 95 & 2.2\\
   & $V-K_S$   & $\left[1.124, 3.010\right]$ &
2.4209 & $-$1.9857 & 0.6720 & $-$0.0819 & 0.0323 & $-$0.0795 & 0.0005 & 102 & 1.0\\
\hline
\end{tabular}
\begin{minipage}{1\textwidth}
$N$ is the number of stars used for the fit after the $3 \sigma$ clipping 
and $\sigma (\%)$ the standard deviation of the percentage differences between 
the measured fluxes and the fluxes calculated from the fitting formula. The
coefficients of the calibrations $b_i$ are given in units of $10^{-5}$ erg 
cm$^{-2}$ s$^{-1}$. 
\end{minipage}
\end{table*}

\begin{figure*}
\begin{center}
\includegraphics[scale=0.74]{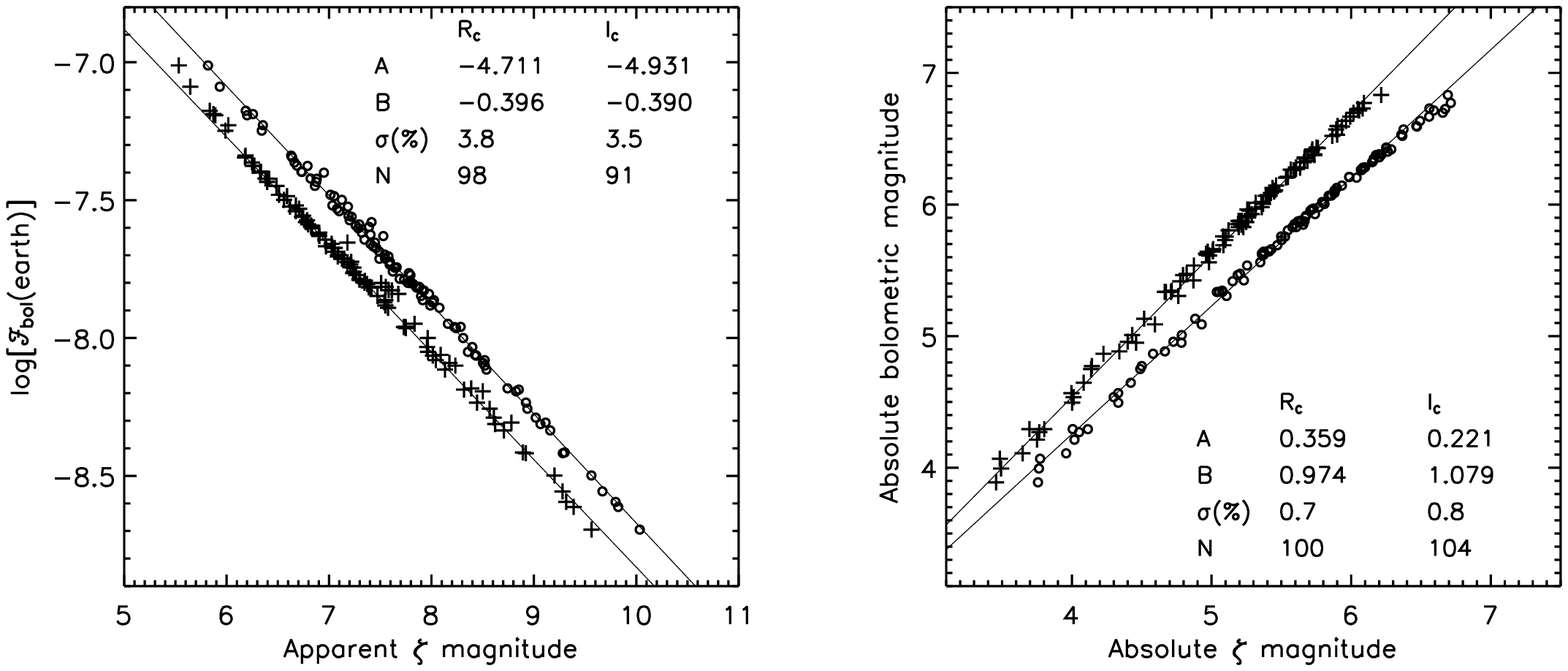}
\caption{{Left panel: relation between apparent magnitudes in $\zeta = R_C$ 
(circles) and $\zeta = I_C$ (crosses) band and bolometric fluxes on the Earth 
(in erg cm$^{-2}$ s$^{-1}$). Right panel: relation between absolute magnitudes 
in the same $R_C$ and $I_C$ bands and absolute bolometric magnitudes. 
The fit is in the form 
$\gamma=A+B\zeta$, $\sigma(\%)$ is the standard deviation
of the percentage differences between the measured values and those calculated
from the fit and $N$ is the number of stars employed for the fitting after the
$3 \sigma$ clipping.}}
\label{funny}
\end{center}
\end{figure*}

The scatter in Figure \ref{phi} of the reduced flux in $R_C$ and $I_C$ bands is
only apparently large, due to the small range (in vertical scale) covered by
the reduced fluxes in these two colours (compare also with Figure
\ref{theo}). For the sake of the proposed calibration, the small range covered
by the reduced flux in these two bands allows for a very accurate (at 1\% level
or below) calibration.  Interestingly, for these two bands tight relations also
exist between bolometric fluxes and magnitudes, as seen in Figure \ref{funny}.

\section{The Angular-diameter-colour relation}\label{AngCal}

\begin{figure*}
\begin{center}
\includegraphics[scale=0.65]{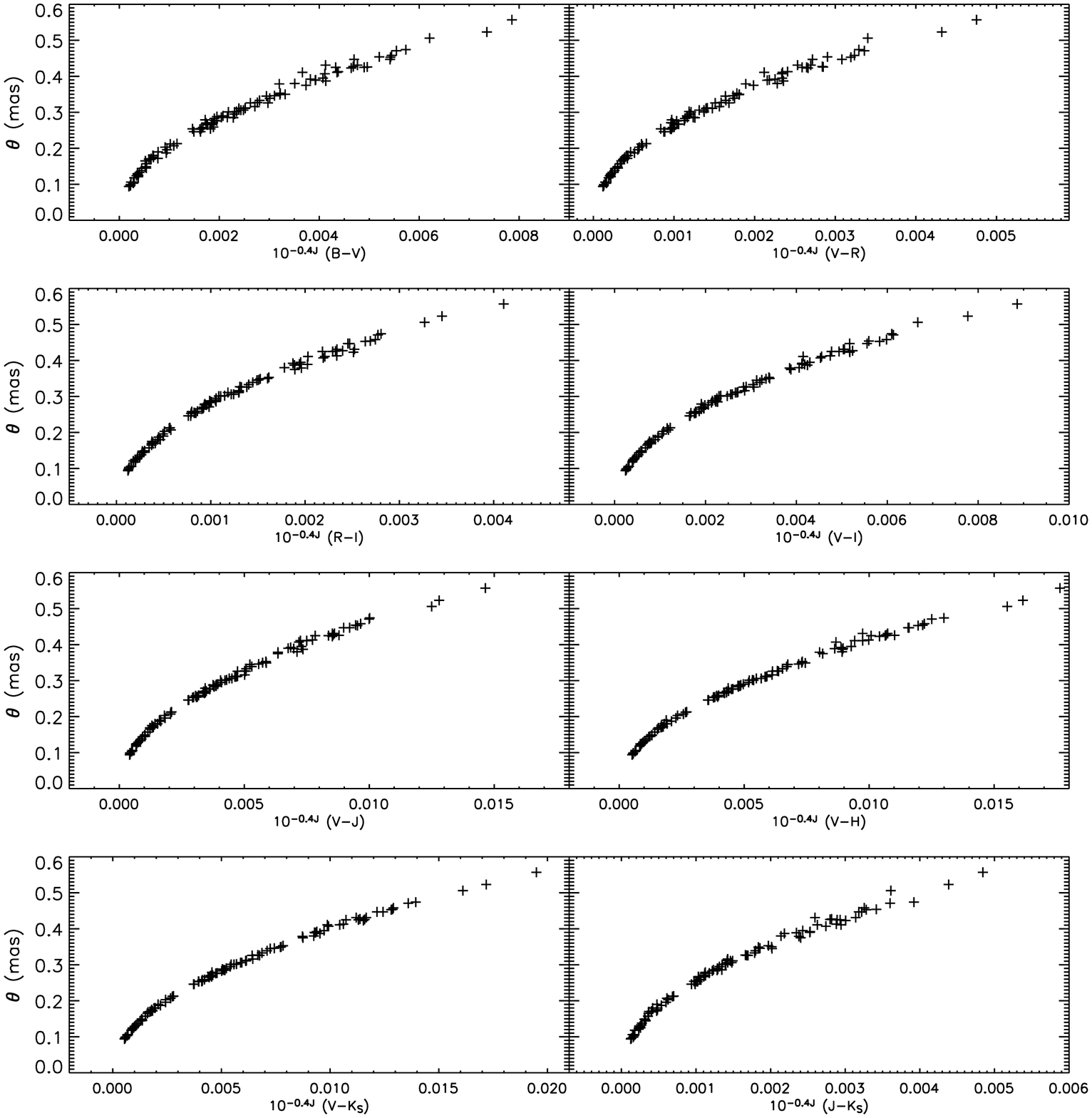}
\caption{{Empirical angular diameter-colour relations for our sample of 
stars.}}
\label{tightheta}
\end{center}
\end{figure*}

Limb-darkened angular diameters can then be readily obtained via equation
(\ref{bolflux}) from the temperature and bolometric flux calibrations given in
Section \ref{irfm} and \ref{BolSec}.  In particular, when using J magnitudes
with various colour indices, we have found very tight and simple relations, as
can be appreciated from Figure \ref{tightheta}. Analogous relations in other
bands are significantly less tight.

We have fitted relations of the form
\begin{equation}
\theta = c_0 + c_1 \sqrt{\phi(m_{J},X)} 
\end{equation}
where
\begin{equation}
\phi(m_{J},X) = 10^{-0.4 m_J} X
\end{equation}
for a given colour index $X$.

\begin{table*}
\centering
\caption{Coefficients and range of applicability of the angular
         diameter calibrations.}\label{angfit}
\begin{tabular}{ccccccc}
Colour & Colour range & $c_0$ & $c_1$ & $N$ & $\sigma(\%)$ & $\sigma$(mas) \\
\hline
$B-V$     & [0.419, 1.270] & 0.00647  & 6.18657 & 102 & 3.4 & 0.008\\
$V-R_C$   & [0.249, 0.755] & 0.00340  & 8.26244 & 103 & 2.7 & 0.008\\
$(R-I)_C$ & [0.268, 0.640] & $-$0.00385 & 8.97758 & 103 & 2.0 & 0.006\\
$V-I_C$   & [0.517, 1.395] & $-$0.00045 & 6.09487 & 102 & 1.6 & 0.005\\
$V-J$     & [0.867, 2.251] & 0.00057  & 4.63147 & 103 & 1.6 & 0.005\\
$V-H$     & [1.016, 2.847] & $-$0.00022 & 4.13607 & 102 & 1.3 & 0.004\\
$V-K_S$   & [1.124, 3.010] & 0.00098  & 4.01335 & 103 & 1.2 & 0.003\\
$J-K_S$   & [0.257, 0.759] & 0.00300  & 7.96076 & 100 & 2.8 & 0.007\\
\hline
\end{tabular}
\begin{minipage}{1\textwidth}
$N$ is the number of stars used for the fit after the $3 \sigma$ clipping.
$\sigma (\%)$ is the standard deviation of the percentage differences between 
the angular diameters obtained via IRFM and those calculated from the fitting 
formula. $\sigma$(mas) is the standard deviation in terms of mas. 
\end{minipage}
\end{table*}

As can be appreciated from Table \ref{angfit}, these relations show remarkably
small scatter (at the few percent level).  In particular, the upper range
covered in angular diameters can be used to build a network of small
calibrators for future long-baseline interferometric measurements from readily
available broadband photometry.

The calibration in $J-K_S$ colour is particularly appealing, since in this
photometric index the effect of interstellar extinction is negligible.

The given angular diameter scale can be tested with the G dwarf HD 209458 A for
which the linear radius has been calculated from planetary transit (Brown et
al. 2001). We have taken $V$ magnitude from Hipparcos and $J$, $H$ and $K_S$
from 2MASS. Our predicted value of $\theta=0.227 \pm 0.003$~mas is in excellent
agreement with that of $\theta=0.228 \pm 0.004$~mas obtained by Kervella et
al. (2004) using surface brightness relations calibrated by interferometry. The
difference between the two values is only 0.4\% and therefore well below the
possible $\sim 3\%$ offset previously discussed.  Unfortunately, both this
comparison and those in Section \ref{checking} and \ref{rm} have their own
limitations since they are indirect, or sensitive to photometric uncertainties
(for this star, Kervella has to convert the 2MASS $K_S$ magnitude into the
Johnson system used to fit his relations).  When our angular size is translated
into linear radius via Hipparcos parallaxes we obtain
$R=1.150\,\pm\,0.056\,R_{\odot}$ slightly smaller than that of Kervella et
al. (2004) but in better agreement with the direct estimate of
$R=1.146\,\pm\,0.050\,R_{\odot}$ obtained by Brown et al. (2001) via $HST$
time-series photometry (although all determinations agree well within the error
budget).  Unfortunately, in this case the uncertainty in parallax dominates the
bulk of the $\sim5\%$ uncertainty on the radius.

\begin{table*}
\centering
\caption{Predicted angular diameter for HD 209458 A.}\label{hd209}
\begin{tabular}{cccccccc}
$V$ & $J$ & $H$ & $K_S$ & $\theta_{V-J}$ & $\theta_{V-H}$ & $\theta_{V-K_{S}}$ &
  $\theta$\\ 
\hline
7.65 & 6.591 & 6.366 & 6.308 & 0.230 & 0.225 & 0.224 & 0.227\\
$\pm0.01$ & $\pm 0.020$ & $\pm 0.038$ & $\pm 0.026$ &  &  & & $\pm 0.003$\\
\hline
\end{tabular}
\begin{minipage}{1\textwidth}
$\theta_{\xi-\zeta}$ are the angular diameters obtained when $(V-J)$,
  $(V-H)$ and $(V-K_S)$ colour indices are used. Since the accuracy of the
  calibration in these bands is very similar (see Table \ref{angfit}), 
  the final angular diameter is the average of the three values weighted by the
  photometric errors.  
\end{minipage}
\end{table*}

\section{Conclusions}

We have used the Infrared Flux Method to deduce fundamental stellar parameters
of effective temperature, bolometric magnitude and angular diameter for a
sample of 104 G and K dwarfs. This semi-direct method is mostly based on
empirical data obtained from our own observations or carefully selected from
the literature in order to achieve the highest accuracy feasible. Our stars all
have excellent $BV(RI)_C JHK_S$ photometry, excellent parallaxes and good
metallicities.

Most of the bolometric flux of our stars is seen in the optical and infrared
bands covered $BV(RI)_C JHK_S$. For the remaining 15 to 30\% of the flux lying
outside these bands, model atmospheres have been used. 
Good to very good agreement is found for the colours of the
stars and the synthetically derived colours of a number of sets of model
atmospheres in the literature (ATLAS9, MARCS, BaSel 3.1). 

The zero-points of our temperature, luminosity and angular diameter scales
(Sections \ref{irfm}, \ref{BolSec} and \ref{AngCal}) depend only on our
adopted absolute calibration of Vega. 
The advantages and disadvantages of our use of Vega are discussed in Appendix
A. The main advantage is that Vega's absolute calibration has been extensively
studied over a wide wavelength range via satellite measurements. This allows us
to put firm constraints on the systematic uncertainties. A likely disadvantage
is that Vega is known to be pole-on and a fast rotator which prevents the use
of a unique Vega model atmosphere in fitting the observations.

Our calibration of effective temperature is found to be some 100~K hotter than
similar determinations in the literature, while it is also found to be in close
agreement with effective temperature scales based on spectroscopic methods. To
reduce our temperature scale by 100 K, the required changes to Vega's 2MASS
zero-points and/or absolute calibration are significantly larger than the known
uncertainties in these quantities. Neither can the model atmospheres be blamed,
because we find such good agreement between the model atmosphere colours and
the empirical data.  We worked hard to impute inconsistency to one or other of
the scales, but were unable to resolve the issue beyond any reasonable doubt. 
Our angular diameters are smaller by about 3\% with respect to other indirect 
determinations, but they seem more in line with the predictions of 
asteroseismology and 3D model atmospheres.
We conclude that temperatures, luminosities and angular diameters calibrations 
for lower main sequence G and K dwarfs retain systematics of the order of a 
few percent.

The high quality and homogeneity of the data produce very tight empirical
colour--metallicity--temperature and angular diameter--colour calibrations. In
particular, the relation between angular diameters and magnitudes in the $J$
band is remarkable and it indicates an high sensitivity of this band to angular
diameters.
Since many lower main sequence stars are known to host planets, our relations
can be used to accurately determine the physical properties of the parent star
thus allowing to effectively break the degeneracy between the properties of the
stars and that of the planets when extremely high precision data are
unavailable (e.g. Bakos et al. 2005, Sato et al. 2005).

Future interferometric measurements of angular diameters would go a long way to
addressing the uncertainties we have shown as well as to test our findings. In
particular, the shape of the visibility function in the second lobe would
directly probe limb-darkening corrections (e.g. Bigot et al. 2006) for our
stars. 
Likewise, tying the absolute calibration and the observed colours to a solar 
twin would probably reduce many of these uncertainties. At present good 
candidate
solar twins are HD 146233 (Porto de Mello \& da Silva) and HD 98618 (Mel\'endez
et al. 2006). Accurate interferometric measurements together with high
precision photometry for a set of nearby solar-like stars would permit to set
the absolute calibration via solar analogs directly (e.g. Campins et al. 1985).
Besides, extremely high precision multiple bandpass studies of transiting 
extra-solar
planets would provide direct angular diameter estimates for our stars, and will
further test our temperature scale.

\section*{Acknowledgments}

We are very grateful to Andrei Berdyugin for instruction in the use of the
remotely operated telescope at La Palma. We thank Steve Willner for very useful
discussions on the absolute calibrations and Johan Holmberg and Jorge
Mel\'endez for many constructive comments and a careful reading of the
manuscript, as well as the referee for the same kindness. LC and CF acknowledge
the hospitality of the Research School of Astronomy and Astrophysics at Mount
Stromlo where part of this work was carried out. We are very grateful to the
Academy of Finland for considerable financial support (grants nr.~206055
and~208792). LC acknowledges the support of the Magnus Ehrnrooth Foundation
and a CIMO fellowship. LP further acknowledges the support of a EU Marie Curie
Intra-European Fellowship under contract MEIF-CT-2005-010884. This research has
made use of the SIMBAD database, operated at CDS, Strasbourg, France. This
publication makes use of data products from the Two Micron All Sky Survey,
which is a joint project of the University of Massachusetts and the Infrared
Processing and Analysis Center/California Institute of Technology, funded by
the National Aeronautics and Space Administration and the National Science
Foundation.

\appendix

\section[]{Broad-band synthetic photometry and absolute calibration}

The choice of the zero-points is of critical importance to set synthetic and
real photometry on the same scale.  Historically, the zero-points of the $UBV$
system were defined ``in terms of unreddened main-sequence stars of class A0
\ldots with an accuracy sufficient to permit the placement of the zero-point to
about 0.01 mag'' (Johnson \& Morgan, 1953).  The pioneering work of Johnson was
continued and extended to other bands by Johnson himself over the years; in the
southern hemisphere Cousins (1976, 1978, 1980) and astronomers at the South
African Astronomical Observatory (SAAO) refined its accuracy and colours range
(Menzies et al. 1989; Kilkenny et al. 1998).  Similar work was carried out in
the northern hemisphere by Landolt (1992) at the CTIO. Subtle differences exist
between different works (see Bessell 2005 for a review) and especially between
the SAAO and the CTIO system (Bessell 1995).  Nevertheless, the $BV(RI)_{C}$
system is nowadays a well defined one whose main characteristics and bandpasses
are given in Bessell (1990b).

The basic equation of synthetic photometry\footnote {The formulation given is
suited for the energy integration that characterize traditional systems based
on energy-amplifier devices like the Johnson-Cousins. Nevertheless, since the
2MASS transmission curves are already multiplied by $\lambda$ and renormalized
(Cohen et al. 2003), eq. (\ref{synth}) can also be used for the 2MASS
photo-counting integration. Notice that even if our $BV(RI)_{C}$ measurements
have been done with a photo-counting device (CCD), we are placing our
photometry on a system of standard stars defined with the use of
photomultiplier and therefore energy integration is the most appropriate.} in a
given band $\zeta$ reads as follows (e.g. Girardi et al. 2002):

\begin{equation}\label{synth}
m_{\zeta} = -2.5\log \left[ \left(\frac{R}{d}\right)^{2} 
\frac{\int_{\lambda_{\zeta_{i}}}^{\lambda_{\zeta_{f}}} F(\lambda) T_{\zeta}
(\lambda)\textrm{d}\lambda}{\int_{\lambda_{\zeta_{i}}}^{\lambda_{\zeta_{f}}}
T_{\zeta}(\lambda)\textrm{d}\lambda} \right]+\textrm{ZP}_{\zeta}
\end{equation}

where $F(\lambda)$ is the flux of the synthetic spectrum (given in erg
cm$^{-2}$ s$^{-1}$ \AA$^{-1}$), $T_{\zeta}(\lambda)$ the transmission curve of
the filter comprised in the interval
$(\lambda_{\zeta_{i}},\lambda_{\zeta_{f}})$ and $\textrm{ZP}_{\zeta}$ the
zero-point for the $\zeta$ band. The ratio between the radius $R$ of the star
and its distance $d$ from us is known as the dilution factor and relates to the
angular diameter $\theta$ (corrected for limb darkening) via:

\begin{equation}
\theta=2\,\frac{R}{d}
\end{equation} 

where $\theta$ is in radians.  We also note that when all quantities in the
logarithmic term of eq.  (\ref{synth}) are known that gives the absolute
calibration in $\zeta$-band (see Table \ref{absflux}) for a star with spectrum
$F(\lambda)$.

While the zero-points of observational photometry are defined on an ensemble of
well measured stars, for synthetic photometry a common choice is to set the
zero-points using a reference star for which all the physical quantities in
eq. (\ref{synth}), i.e. the spectrum $F(\lambda)$ (usually synthetic), the
dilution factor and the observed apparent magnitudes or colour indices are
known in detail.

This star is usually Vega, for which we have used the magnitudes and colours
given in Table \ref{vzpo}. The optical colours are from Bessell (1990a), 
whereas for the 2MASS system they come from Cohen et al. (2003).
The latter were determined post-facto, comparing the observed 2MASS 
magnitudes of 33 stars with the values predicted from the absolutely 
calibrated templates built by Cohen et al. (2003) 
within the ``Cohen-Walker-Witteborn'' (1992a) framework.

\begin{table*}\label{vzpo}
\centering
\caption{Observed magnitudes for Vega.}\label{vzpo}
\begin{tabular}{cccccccr}
\hline
 $B$ & $V$  & $R_{C}$ & $I_{C}$ &  $J$   &  $H$   &  $K_{S}$ & Ref.\\
\hline
0.02 & 0.03 & 0.039  & 0.035   &        &        &       & Bessell 1990a\\
     &      &        &    & $-$0.001 & +0.019 & $-$0.017 & Cohen et al. 2003\\
\hline
\end{tabular}
\end{table*}

In principle we could adopt a unique spectrum for Vega and use it to compute
zero-points from $B$ to $K_{S}$; in this way the dependence on the absolute
calibration would cancel out when computing colour indices, since all colours
scale in the same way with the adopted absolute calibration.  In practice we
use a slightly different approach using two models, one for the absolute
calibration in the optical and one in the infrared -- each consistent with the
corresponding data source.

The 2MASS magnitudes for Vega are deduced in the absolutely calibrated system
that is ultimately defined on the Kurucz spectra for Vega and Sirius used by
Cohen et al. (1992b). Therefore, for the $JHK_{S}$ bands we have adopted the
zero-magnitude fluxes provided by Cohen et al.  (2003).  The zero-magnitude
fluxes proposed by Cohen for Vega and eight of the primary and secondary stars
in the calibration network of ``Cohen-Walker-Witteborn'' has been recently
confirmed in the $4$--$24\, \mu m$ range by the \emph{Midcourse Space
Experiment (MSX)} flux calibration (Price et al. 2004) to be accurate around
1\% and thus well within the global error of 1.46\% quoted by Cohen, even
though it seems that the fluxes of Cohen in the infrared should be brightened
by 1\%. Also, Tokunaga \& Vacca (2005) have shown the Vega absolute
calibration of Cohen et al. (1992b) and the model--independent measurements of
Megessier (1995) to be identical within the uncertainties in the range $\sim$
1--4 $\mu m$.

For the absolute calibration in the optical bands we have adopted a Kurucz
(2003) synthetic spectrum for Vega with $T_{\mathrm eff}=9550$ K,
$\log(g)=3.95$, [M/H] = $-$0.5 and microturbulent velocity $\xi = 2\,\kms$. The
resolving power used has been 500. This model has recently shown excellent
agreement with the \emph{Space Telescope Imaging Spectrograph (STIS)} flux
distribution over the range $0.17$--$1.01\, \mu m$ (Bohlin \& Gilliland 2004).
In particular, in the regions of the Balmer and Paschen lines the \emph{STIS}
equivalent widths differ from the pioneering work of Hayes (1985) (and used by
Colina et al. 1996 to assemble a composite spectrum of Vega) but do agree with
the predictions of the Kurucz model.

Once the synthetic spectrum of Vega in the optical bands is chosen, we need to
scale it in order to match the absolute flux of Vega as measured on the Earth
at a certain fixed wavelength ($\lambda$). The monochromatic flux given by the
model $F_{\textrm{\scriptsize{model}}}(\lambda)$ is simply related to the same
monochromatic flux as measured on the Earth
$F_{\textrm{\scriptsize{Earth}}}(\lambda)$ by

\begin{equation}\label{SP}
F_{\textrm{\scriptsize{Earth}}}(\lambda)=
\left(\frac{R}{d}\right)^2F_{\textrm{\scriptsize{model}}}(\lambda) 
\end{equation}

where the dilution factor $(R/d)^2$ is the ratio between the radius of Vega and
its distance. In principle the dilution factor can be deduced from direct
measures of Vega's angular diameter. In practice, since such measures are more
uncertain than direct measures of the flux, we proceed the other way
around. Taking the flux value at 5556 \AA$ $ from Megessier (1995)
$F_{\textrm{\scriptsize{Earth}}}(5556\,\textrm{\AA})=3.46 \times 10^{-9}$ erg
cm$^{-2}$ s$^{-1}$ \AA$^{-1}$ and at the same wavelength from the Kurucz (2003)
Vega model $F_{\textrm{\scriptsize{model}}}(5556\,\textrm{\AA})= 5.5015572
\times 10^7$ erg cm$^{-2}$ s$^{-1}$ \AA$^{-1}$ we obtain $(R/d)^2=6.2891286
\times 10^{-17}$. This value implies and angular diameter of 3.272 mas for
Vega, which compares very well with the interferometric angular diameter
measurements of $3.24\pm0.07$ mas (Code et al. 1976), $3.28\pm0.06$ mas (Ciardi
et al. 2001) and $3.225\pm0.032$ (Mozurkewich et al. 2003).

We chose to use the Megessier flux in accordance with the absolute scale
adopted for the STIS (Bohlin \& Gilliland 2004) rather than the value of $3.44
\times 10^{-9}$ erg cm$^{-2}$ s$^{-1}$ \AA$^{-1}$ $\pm 1.45$\% found in Hayes
(1985) and used by Cohen to tie his absolute calibration in the infrared. A
thorough discussion of the differences in the infrared is given by Bohlin \&
Gilliland (2004); here we just mention that this choice together with the
Kurucz (2003) model in the infrared would give a flux density of Vega about 2\%
lower than in Cohen, thus worsening the agreement with \emph{MSX}.  On the
other hand, the \emph{STIS} absolute calibration in the optical is expected to
be better than 2\% (Bohlin 2000). In this range, the increase of $0.6$\%
relative to Hayes translates in absolute fluxes that are higher by the same
factor and in differences for the $BV(RI)_{C}$ zero-points of 0.006 mag.

\begin{table*}\label{absflux}
\centering
\caption{Absolute calibration and effective wavelength of the ground-based 
optical-IR photometry of Vega used in this work. Quantities tabulated 
correspond to the definition of the zero magnitude in each filter.}\label{absflux}
\begin{tabular}{ccccc}
\hline
Band & $\lambda_{\mathrm eff}$ & Monochromatic Absolute Flux & Uncertainty & Ref.\\
\hline
 & \AA & erg cm$^{-2}$ s$^{-1}$\AA$^{-1}$ & erg cm$^{-2}$ s$^{-1}$\AA$^{-1}$ & \\
\hline
$B$  &  4362   &  6.310E-9	&  6.310E-11 & This paper \\
$V$  &	5446   &  3.607E-9	&  3.607E-11 & This paper \\
$R_C$ &	6413   &  2.153E-9	&  2.153E-11 & This paper \\
$I_C$ &	7978   &  1.119E-9	&  1.119E-11 & This paper \\
$J$  &	12285  &  3.129E-10     &  5.464E-12 & Cohen et al. (2003) \\
$H$  &	16385  &  1.133E-10     &  2.212E-12 & Cohen et al. (2003) \\
$K_S$ &	21521  &  4.283E-11     &  8.053E-13 & Cohen et al. (2003) \\
\hline
\end{tabular}
\begin{minipage}{1\textwidth}
The effective wavelengths associated to each filter are computed in 
accordance with Appendix B. The error in $BV(RI)_C$ bands are computed 
assuming an arbitrary uncertainty of 1\% to the absolute flux. Lacking the 
actual uncertainties in the measurement of the filters' transmission curve, 
the adopted uncertainty is consistent with a wavelength--independent filter 
uncertainty in addition to the 0.7\% absolute uncertainty at 5556 \AA{} from 
Megessier (1995).
\end{minipage}
\end{table*}

Recent work by Gulliver et al. (1994), Peterson et al. (2004, 2006) and
Aufdenberg et al. (2006) indicate that Vega is pole-on and fast rotator. Thus
standard model atmospheres are not appropriate for Vega, as discussed by Bohlin
\& Gilliland (2004). Nonetheless \emph{STIS} in the optical and \emph{MSX} in
the infrared have confirmed the adopted absolute calibrations to be accurate at
the percent level. It is interesting to note that Peterson et al. (2006) give
an estimate of the shift in Vega's zero-points as a consequence of the higher
infrared flux due to its rapidly rotating nature. Though only indicative, as
the authors say, the shifts for their latest non-rotating synthetic spectrum in
the infrared amount to $0.05-0.07$ mag. The net effect of such shifts would be
to reduce the temperatures recovered via IRFM. Considering that in the infrared
we are adopting the absolute calibration given by Cohen (already some $\sim
2\%$ brighter as compared to the latest Kurucz model) this reduces the size of
the required shifts to about $0.02-0.04$ mag. The agreement within 1\% with
\emph{MSX} would however require a shift of only $0.01$ mag. Since the proposed
shifts are only indicative and the measurements confirm the adopted calibration
within the errors, we adopt the pragmatic choice of using the model that at any
given range better agrees with the data.  It is evident from the above
discussion that there is no consistent published atmospheric model for Vega at
both visible and infrared wavelengths and this justifies our adoption of the
Kurucz (2003) or Cohen models at different wavelengths. Likewise, the use of
composite model atmospheres for Vega is also advocated from Peterson et
al. (2006).  For synthetic photometry the use of composite spectra is often
avoided on the reason that fixing the zero-points, possible physical
deficiencies in the synthetic spectrum of Vega are likely to be present in the
synthetic library and therefore they would cancel out (e.g. Girardi et
al. 2002).  In our case the situation is not so clear, since synthetic spectra
give now excellent comparison with the real ones and the deficiencies for Vega
are likely to be due to the particular nature of this star.

Sirius is often used as an additional fundamental colour standard
(e.g. Bessell, Castelli \& Plez 1998), allowing one to control the possible
problems of putative variability and an IR excess for Vega. First seen by
Aumann (1984) beyond 12 $\mu m$, Vega's variability at shorter wavelengths has
long been a matter of discussion (see e.g. Bessell, Castelli \& Plez 1998;
Ciardi et al. 2001), though the rapidly rotating model could actually resolve
the controversy (Peterson at al. 2006).  The most recent data from \emph{MSX}
(Price et al. 2004) and \emph{SPITZER} (Su et al. 2005) support models where
Vega's infrared excess and variability is due to a cold dust disk. The IR
excess increase steadily longward of $12\,\mu m$ whereas at $4\,\mu m$ the
differences between model atmosphere and measurements are entirely consistent
within the uncertainties and therefore should not affect the optical and
near-infrared region (0.36 -- 2.4 $\mu m$) we are working with.  For the sake
of completeness we have tested the differences in the derived $BV(RI)_{C}$
zero-points when including Sirius in the calibration.  Adopting the latest
Kurucz model available for it and the colours from Bessell (1990a), the
differences in the derived zero-points between the use of Vega only or Vega
plus Sirius range from 0.008 mag in $B-V$ to 0.002 mag in $(R-I)_{C}$. Such
differences are below the uncertainties in magnitude and colours.  In addition,
since absolute calibration and magnitudes for Sirius in the 2MASS system are
not available, we have decided to adopt only Vega as our standard star. This
also makes clearer the comparison between synthetic colours and the IRFM that
strongly depends on the Vega absolute calibration.

The transmission curves used come from Bessell (1990b) for the Johnson-Cousins
$BV(RI)_C$ system and from Cohen et al. (2003) for the 2MASS $JHK_{S}$ system.

The 2MASS transmission curves carefully incorporate the effect of the optical
system, the detector quantum efficiency and the site-specific atmospheric
transmissions. The effect of the telescope optics on the estimated $BV(RI)_C$
passband are not counted, but in the $V$ band this has been proved to cause
uncertainty below few millimag (Colina \& Bohlin, 1994), and thus smaller than
any possible observational error we are going to compare with.

We note that all our photometry is reduced to zero airmass and when we refer to
measurements on the Earth, we always mean at the top of the Earth's atmosphere.

\section[]{Characteristic Parameters of a Photometric System}

In the present work we have reduced broadband (heterochromatic) measurement in
a band $\zeta$ to a monochromatic flux ($\mathcal{F_{\zeta}}$) by means of the
following relation:

\begin{equation}
\mathcal{F_{\zeta}}=\frac{\int_{\lambda_{\zeta_i}}^{\lambda_{\zeta_f}} 
F(\lambda) T_{\zeta}(\lambda) \textrm{d}\lambda}{\int_{\lambda_{\zeta_i}}^
{\lambda_{\zeta_f}} T_{\zeta}(\lambda) \textrm{d}\lambda},
\end{equation}

where the integration is done in the interval $\lambda_{\zeta_i},
\lambda_{\zeta_f}$ that comprise a given passband $T_{\zeta}(\lambda)$.  We
remark that the above formulation is suited for both energy integration in the
optical and photo-counting in the IR, provided that we are dealing with the
Bessell (1990b) and Cohen et al. (2003) transmission curves (see Appendix A).

Assuming that the function $F(\lambda)$ is continuous and that
$T_{\zeta}(\lambda)$ does not change sign in the interval $\lambda_{\zeta_i},
\lambda_{\zeta_f}$, the generalization of the mean value theorem states that
there is at least one value of $\lambda$ in the interval $\lambda_{\zeta_i},
\lambda_{\zeta_f}$ such that

\begin{equation}
F(\lambda_i)\int_{\lambda_{\zeta_i}}^{\lambda_{\zeta_f}}T_{\zeta}(\lambda)
\textrm{d}\lambda=\int_{\lambda_{\zeta_i}}^{\lambda_{\zeta_f}}F(\lambda)
T_{\zeta}(\lambda)\textrm{d}\lambda.
\end{equation}

Rearranging this, we obtain

\begin{equation}
\mathcal{F}_{\zeta}=F(\lambda_i)=\frac{\int_{\lambda_{\zeta_i}}^
{\lambda_{\zeta_f}}F(\lambda) T_{\zeta}(\lambda) \textrm{d}\lambda}
{\int_{\lambda_{\zeta_i}}^{\lambda_{\zeta_f}} T_{\zeta}(\lambda) 
\textrm{d}\lambda},
\end{equation}

where $\lambda_i$ is the \emph{isophotal wavelength}. The isophotal wavelength
is thus the wavelength which must be given to the monochromatic quantity
$F(\lambda_i)$ obtained from a heterochromatic measurement.

Stellar spectra do not necessarily satisfy the requirements of the mean value
theorem for integration, as they exhibit discontinuities. Although the mean
value of the intrinsic flux is well-defined (and we have extensively used it),
the determination of the isophotal wavelength becomes problematic because
spectra contain absorption lines and hence the definition can yield multiple
solutions.

Several authors avoid using the isophotal wavelength and introduce the
\emph{effective wavelength} defined by the following expression

\begin{equation}
\lambda_{\mathrm eff}=\frac{\int_{\lambda_{\zeta_i}}^{\lambda_{\zeta_f}}\lambda
F(\lambda)T_{\zeta}(\lambda)\textrm{d}\lambda}{\int_{\lambda_{\zeta_i}}^
{\lambda_{\zeta_f}}F(\lambda)T_{\zeta}(\lambda)\textrm{d}\lambda}.
\end{equation} 

This is the mean wavelength of the passband as weighted by the energy
distribution of the source over the band. The effective wavelength is thus an
approximation for $\lambda_i$, however since the monochromatic magnitude at the
effective wavelength is almost identical to that at the isophotal wavelength
(Golay 1974), we have used the effective wavelength through all our work.
 
We have verified that the detailed choice of the wavelength associated to a
monochromatic flux is not crucial for the derived correction of the
$\mathcal{C}$ factor (eq. \ref{cfact}).

When dealing with photometric data, a way to obtain monochromatic flux from
heterochromatic measurements is provided by the $q$ factor.  Ideally the
$q$-factor (eq. \ref{qmono}) should be determined from spectroscopic data but
in practice we rely on a grid of models. Since the absolute monochromatic flux
for a given band is determined as described in Appendix A, the definition of
the $q$-factor for a star with spectra $F(\lambda)$ is now:

\begin{equation}
q(\lambda_{\textrm{\tiny{IR}}})=\frac{\int_{\lambda_{i}}^{\lambda_{f}} 
T_{\textrm{\tiny{IR}}}(\lambda)\textrm{d}\lambda}{\frac{1}
{F(\lambda_{\textrm{\tiny{IR}}})}\int_{\lambda_{i}}^{\lambda_{f}} F(\lambda)
T_{\textrm{\tiny{IR}}}\textrm{d}\lambda}
\end{equation}

This definition slightly differs from that given by Alonso et al. 
(1994, 1996b) to
reflect the different absolute calibration we have adopted (i.e. argument of
the logarithm in equation \ref{synth}).  However, the different definition of
the $q$ factor has only minor effects on the recovered stellar
parameters. Using the definition of Alonso et al. (1996b) would go in the
direction of making temperatures hotter by 20 to 30~K on average, the
difference increasing with increasing temperature.  On average also
luminosities would be brighter by 0.14\% and angular diameters smaller by
0.9\%.

\end{document}